\documentclass{aa}  

\usepackage{siunitx}
\usepackage{ulem}
\usepackage{mathtools}
\usepackage{enumitem}
\usepackage[utf8]{inputenc}
\usepackage{tikz,xcolor,hyperref}

\usepackage{graphicx}
\usepackage{txfonts}
\usepackage{lipsum}
\usepackage{subcaption}         % necessary for continued figures, example in section 3
\usepackage{lscape}             % to rotate a single page table, example in appendix.
\usepackage{placeins}           % useful with \FloatBarrier, to keep 
\usepackage{color}

\newcommand{\LedaReal}[1]{{Leda~1044720}}
\newcommand{\Leda}[1]{{W2246f}}

\definecolor{lime}{HTML}{A6CE39}

\begin{document}

   \title{A Deep Study of the Spiral Galaxy \Leda\ }

   %\subtitle{Subtitle}

%%%%%%%%%%%%%%%%%%%%%%%%%%%%%%%%%%%%%%%%
% Please do not include ORCIDs next to author names.
% Only ORCIDs authenticated by individual authors in EDP Sciences editorial system will be taken into account.
% ORCIDs included here will be removed.
%%%%%%%%%%%%%%%%%%%%%%%%%%%%%%%%%%%%%%%%

   \author{Evelyn J. Johnston\inst{1} \fnmsep \corrauth{evelyn.johnston@mail.udp.cl} % use \corrauth for the corresponding author
        \and Sorya Lambert\inst{1}
        \and Aparna Nair\inst{2}
        \and Alejandra Z. Lugo-Aranda\inst{3}
        \and Manuel Aravena\inst{1,4}
        \and Roberto J. Assef\inst{1}
        \and Tanio Diaz Santos\inst{5,6}
        \and Román Fernández Aranda\inst{7}
        \and Hyunsung D. Jun\inst{8}
        \and Guodong Li\inst{9}
        \and Mai Liao\inst{1}
        \and Devika Shobhana\inst{1}
        \and Chao-Wei Tsai\inst{10,11,12} }

 % \author{Evelyn J. Johnston\inst{1}\fnmsep,
 %        Sorya Lambert\inst{1},
 %        Aparna Nair\inst{2},
 %        Alejandra Z. Lugo-Aranda\inst{3},
 %        Manuel Aravena\inst{1,4},
 %        Roberto J. Assef\inst{1},
 %        Tanio Diaz Santos\inst{5,6},
 %        Román Fernández Aranda\inst{7},
 %        Hyunsung D. Jun\inst{8},
 %        Guodong Li\inst{9},
 %        Mai Liao\inst{1},
 %        Devika Shobhana\inst{1},
 %        Chao-Wei Tsai\inst{10,11,12},
 % }

 \institute{Instituto de Estudios Astrof\'isicos, Facultad de Ingenier\'ia y Ciencias, Universidad Diego Portales, Av. Ej\'ercito Libertador 441, Santiago, Chile. 
            % \email{evelyn.johnston@mail.udp.cl}
       \and
           Instituto de Astronom\'ia y Ciencias Planetarias, Universidad de Atacama, Avenida Copayapu 485, 1530000, Copiap\'o, Chile %Aparna
       \and
           Universidad Nacional Autónoma de México, Instituto de Astronomía, AP 106, Ensenada 22800, BC, México %Alejandra
       \and
           Millenium Nucleus for Galaxies (MINGAL) %Manuel2
       \and
           Institute of Astrophysics, Foundation for Research and Technology Hellas (FORTH), Heraklion, 70013, Greece %Tanio1
        \and
           School of Sciences, European University Cyprus, Diogenes Street, Engomi 1516, Nicosia, Cyprus %Tanio2
       \and
           Centro de Astrobiolog\'ia (CAB), CSIC-INTA, Carretera de Ajalvir km 4, Torrej\'on de Ardoz 28850, Madrid, Spain %Roman
       \and
           Department of Physics, Northwestern College, 101 7th St SW, Orange City, IA 51041, USA %Hyunsung
       \and
           Kavli Institute for Astronomy and Astrophysics, Peking University, Beijing 100871, People's Republic of China %Guodong
       \and
           National Astronomical Observatories, Chinese Academy of Sciences, 20A Datun Road, Beijing 100101, China %Chao-Wei
       \and
           Institute for Frontiers in Astronomy and Astrophysics, Beijing Normal University,  Beijing 102206, China %Chao-Wei
       \and
           School of Astronomy and Space Science, University of Chinese Academy of Sciences, Beijing 100049, China %Chao-Wei
           }

   \date{Received September 30, 20XX}

\titlerunning{A Deep Study of the Spiral Galaxy \Leda\ }
\authorrunning{Johnston et al}

% \abstract{}{}{}{}{}
% 5 {} token are mandatory
 
  \abstract
   {}
   {In this era of large surveys and statistical studies of galaxies, the beauty in the details of individual galaxies is often lost. In this paper we present a deep study of the spiral galaxy \Leda\. with MUSE, exploring the spatially-resolved stellar and ionized gas properties to understand how it formed and evolved over time. The unusually deep observations of this galaxy with MUSE give us a rare opportunity to study this phenomenon with better spatial resolution than can normally be achieved with the current IFU surveys of galaxies at a similar redshift ($z\sim0.09$).} 
   %{We analyse the stellar and gas kinematics, and the spatially resolved stellar populations and  gas properties, including the gas metallicity and the dominant ionization sources.}
   {We analyse the stellar and gas kinematics, as well as the spatially resolved stellar populations and gas properties, including gas metallicity and the dominant ionization sources. The derived properties include the stellar mass, radial profiles of luminosity- and mass-weighted mean ages and metallicities, and ionized gas characteristics such as E(B$-$V), H$\alpha$ extinction, dust-corrected H$\alpha$ flux, oxygen abundance using the O3N2 calibrator, H$\alpha$ luminosity, and H$\alpha$-based star formation rate.}
   {Analysis of the stellar populations  revealed a negative metallicity gradient, and the mass-weighted ages showed uniformly flat ages across the disc while the luminosity-weighted ages show a negative gradient. We find that the gas metallicity and star formation rate density also drop in the central region of the galaxy where the older luminosity-weighted stellar populations are found. Analysis of the WHAN and WHaD diagrams reveal that in fact the central region is retired while the rest of the disc is star forming.}
   {We conclude that \Leda\. is a nice example of a central low-ionization emission-line region (cLIER) galaxy, where the central kpc is dominated by old, metal-poor stars with little star formation. The central LIER emission is primarily powered by hot evolved stars, while the rest of the disc  displays ongoing star formation. These findings are consistent with a scenario of inside-out quenching. }


   \maketitle
\nolinenumbers

%%%%%%%%%%%%%%%%%%%%%%%%%%%%%%%%%%%%%%%%%%%%%%%%%%%%%%%%%%%%%%
\section{Introduction}

As astronomical research makes increasing use of large surveys and statistical samples, there is still a lot to learn from studying the fine details of individual galaxies to better understand these processes. Spiral galaxies are home to the ongoing recycling of gas in the Universe, with clouds of gas collapsing to trigger the formation of new stars while the death of older stars injects new metals, such as, oxygen, nitrogen, sulphur, back into the ISM where it can enrich the next generation of stars to be created \citep{Matteucci_1989}. Over the last decade, integral field unit (IFU) spectroscopy has revolutionized the study of the structural components of all types of galaxies, including spirals, providing spatially resolved spectroscopic information across their entire extent. This information has shed new light on how galaxies have formed and evolved compared to what could be learnt previously from multi-object or long-slit spectroscopy alone.

For example, long-slit spectroscopy found clear gradients in the stellar ages and metallicities across galaxy discs, where generally the older and more metal-rich stellar populations lie in inner regions of galaxies, suggesting an inside-out growth over time \citep{Tissera_2016, Goddard_2017, Peterken_2020, Lara_2022, Pessa_2023}. However, IFU spectroscopy has revealed that these gradients are not smooth, but instead show bumps or flatter regions, particularly close to spiral arms and bars. Such features indicate that these structures are driving the migration of gas and stars,  meaning that the older stars may now lie very far from where they were formed \citep{DiMatteo_2013, Ruiz_2017}. Thus, it has become clear that spiral galaxies continue to evolve and change throughout their lifetimes.

IFU spectroscopy also allows us to map out the stellar and gas kinematics of spiral galaxies. While in most cases these kinematics maps will show smooth variations across the discs, perturbations can indicate recent interactions or mergers. For example, it is thought that the presence of kinematically decoupled cores (KDCs) and counter-rotating discs indicate the remnant of a smaller galaxy that has merged with the spiral galaxy. During this merger, the infalling galaxy has lost the majority of its mass, and the material from its core has settled into the centre of the galaxy and maintains some of the angular momentum of the progenitor galaxy \citep[e.g.][]{Krajnovic_2015,Johnston_2018b, Nedelchev_2019}. The relative sizes of the two kinematic components and the degree of asymmetry can also help us understand the properties of the two progenitor galaxies and the conditions under which they interacted or merged \citep{Bao_2022}. Furthermore, the kinematics maps can also reveal structures such as nuclear discs \citep[e.g.][]{Medling_2014, Gadotti_2020} and AGN jets \citep[e.g.][]{Venturi_2017} that are significantly harder to detect and characterize with long-slit spectra alone. 

Another benefit of IFU spectroscopy is that one can map out the gas properties across a spiral galaxy in order to understand the different ionization sources that dominate in the different regions \citep{Sanchez_2020}. Generally in spiral galaxies, the majority of the disc is dominated by star formation, but in the centres of these galaxies an AGN might be present \citep[e.g.][]{Belfiore_2015, Belfiore_2016, Fensch_2016, Jingzhe_2021}. By mapping out the ratios of different emission lines it is possible to understand the extent and strength of these AGN. For example, \citet{Belfiore_2016} studied a sample of low-ionization nuclear emission-line region galaxies (LINERs) from the MaNGA survey and found that the LINER emission is often not restricted to only the central region of the galaxy, but in some cases can be detected across large parts of the disc. He named these galaxies LIERs to reflect this broader distribution of low ionization emission.

Along with new instruments with large fields of view and high spatial resolution, such as the Multi-Unit Spectroscopic Explorer \citep[MUSE;][]{Bacon_2010} and the Visible Integral-field Replicable Unit Spectrograph \citep[VIRUS;][]{Hill_2021}, came new IFU surveys to observe large samples of galaxies in a more statistical way. For example, the SDSS-IV Mapping of Nearby Galaxies at APO \citep[MaNGA:][]{Bundy_2015} survey observed $\sim10,000$ galaxies within $z<0.15$ out to $1.5-2.5~R_e$, the Calar Alto Legacy Integral Field Area \citep[CALIFA;][]{Sanchez_2016} survey achieved similar coverage for $>600$ galaxies out to $z\sim0.16$, and the Sydney-AAO Multi-object Integral field spectrograph \citep[SAMI;][]{Croom_2012, Bryant_2015} observed a further $\sim3000$ galaxies out to $z\sim0.1$. Together, these three surveys shed new light on galaxy morphology and evolution, and paved the way for smaller, more specific surveys, such as the MUSE Atlas of DISCS \citep[MAD;][]{Erroz_2019}, the Time Inference with MUSE in Extragalactic Rings \citep[TIMER;][]{Gadotti_2019}, and the PHANGS-MUSE survey \citep{Emsellem_2022}. However, one weakness of many of these surveys is that in order to resolve the features within these galaxies, one must observe relatively nearby galaxies which often are larger than the field of view of the instrument. As a result, many of these surveys focus on the properties in the inner regions of the galaxies, and lose the information from the outskirts, where perturbations due to recent interactions and mergers are likely to be seen for longer \citep{Borlaff_2014, Eliche_2018, Martinez_2025}. One new IFU survey that will achieve full coverage of nearby galaxies is the SDSS-V Local Volume Mapper \citep[LVM;][]{Drory_2024}, with a wide field of view of $\sim30\arcmin$, but this coverage comes with the sacrifice of the spatial resolution ($\sim36\arcsec$ fibers).

As a result, despite the abundance of large surveys and statistical studies of galaxies, there is still room for detailed studies of individual galaxies, especially where particularly deep or high resolution data is available. This paper outlines one such study of \Leda\., also known as \LedaReal\., a spiral galaxy located at RA $22^\mathrm{h}46^\mathrm{m}08^\mathrm{s}.3$ and Dec $-05^\circ 26' 24.5''$ (J2000) that happens to lie in the foreground of a deep ($\sim20$~hours exposure time) MUSE datacube studied in Shobhana et al (in prep). The datacube is centered on WISE J224607.6--052634.9 (W2246--0526), a Hot, Dust-Obscured Galaxy \citep[also known as a Hot DOG;][]{Wu_2012} at redshift $z \sim 4.6$ \citep{Diaz_2018}, and the focus of the work by Shobhana et al (in prep) is to study the environment of this galaxy. \Leda\. was previously studied by \citet{Fan_2018} in the context of understanding its contamination on their studies of W2246–0526, and they retrieved a photometric redshift for \Leda\. of $z_{ph} = 0.047$ from the Sloan Digital Sky Survey SkyServer\footnote{http://skyserver.sdss.org/}. The datacube used in this study covers the entire galaxy with high spatial resolution ($\sim0.2$~arcsec pixel$^{-1}$), with the deep exposure time ensuring good signal even in the outermost regions. In this paper we carry out analysis of the stellar and gas kinematics, the stellar populations and the gas properties to better understand the nature of this galaxy.

The paper is laid out as follows: Section~\ref{sec:DR} describes the observations and data reduction; Section~\ref{sec:morphology} describes the visual morphology of the galaxy; Section~\ref{sec:kinematics} explores the stellar and gas kinematics; Section~\ref{sec:stellar_cont} studies the stellar properties; Section~\ref{sec:gas_properties} gives an overview of the gas properties; and Section~\ref{sec:summary} presents our summary and conclusions. Throughout this paper we assume a Hubble constant of $H_0 = 70$~km~s$^{-1}$~Mpc$^{-1}$ \citep{Lelli_2010}, which corresponds to a projected angular scale of 1.83~kpc~arcsec$^{-1}$ and a distance of 377~Mpc  based on the line-of-sight velocity of the galaxy described in Section~\ref{sec:kinematics}.

%Mass is $4.4 \times 10^{10}$ solar masses

%https://ui.adsabs.harvard.edu/abs/2021MNRAS.505.2087K/abstract
%%%%%%%%%%%%%%%%%%%%%%%%%%%%%%%%%%%%%%%%%%%%%%%%%%%%%%%%%%%%%%
\section{Observations and Data Reduction}\label{sec:DR}

The data used in this study were observed with MUSE on the Very Large Telescope (VLT). MUSE is an optical integral-field spectrograph with a field of view of $1\arcmin\!\times\!1\arcmin$, a spatial resolution of 0.2\arcsec/pixel, and a spectral resolving power ranging from $R\simeq1770$ at 4800\,\AA\ to $R\simeq3590$
at 9300\,\AA. The data were observed between July and September 2022 as part of program IDs 106.21F0.001 (PI Diaz-Santos) and 109.2393.001 (PI Assef). The main target for these programs was W2246–0526 \citep{Tsai_2015, Diaz_2016, Diaz_2018, Tsai_2018}, the most luminous obscured quasar in the universe at redshift $z \sim 4.6$  \citep{Fan_2018}. The spiral galaxy studied in this work happens to lie along a similar line of sight and appears in the upper-left quadrant of the FOV of the MUSE datacube. Due to the faintness of the main target, the MUSE datacube is very deep, consisting of 104 exposures each with an exposure time of 675~s, resulting in a total exposure time of 19.5~hours for a single pointing. The data were observed in Wide Field Mode with AO using the nominal wavelength range (WFM-AO-N), and the average seeing was measured to be $\sim0.65$\arcsec\ on the final combined datacube.

\begin{figure*}
\includegraphics[angle=0,width=1\linewidth]{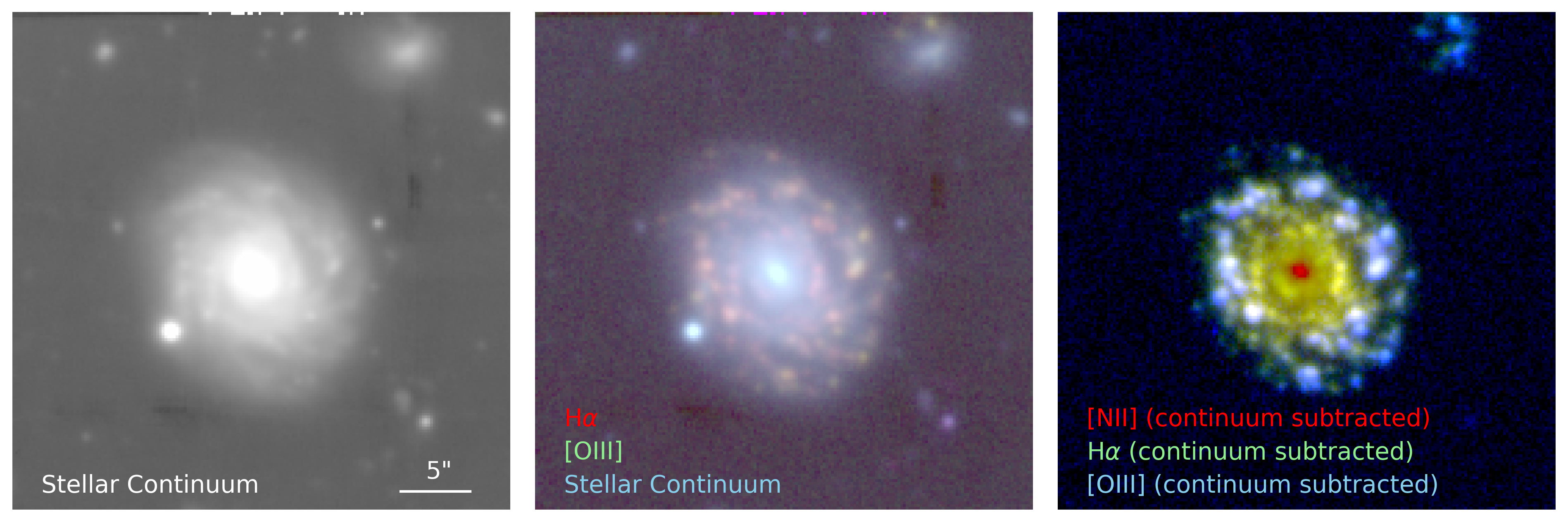}
\caption{Images of \Leda\. and its companion galaxy to the North-West, extracted from the MUSE datacube. On the left is the white light image showing the stellar continuum. In the middle is an RGB image with the H$\alpha$ emission in red, the [\ion{O}{iii}] emission in green and the stellar continuum in blue. In this image the H$\alpha$ and [\ion{O}{iii}] maps have not been continuum subtracted. And on the right is a continuum-subtracted image showing the [\ion{N}{ii}]~$\lambda$6584 emission in red, the H$\alpha$ emission in green and the [\ion{O}{iii}] emission in blue.
%And on the right is a continuum-subtracted image showing just the H$\alpha$ emission in red and the [O\textsc{iii}] emission in green.
The images are orientated with North up and East to the left, and a scale bar is shown in the stellar continuum image for reference, where 5\arcsec\mbox{} corresponds to $\sim 9.15$~kpc.
\label{fig:images}}
\end{figure*}

A full description of the data reduction is given in Shobhana et al (in prep), but in summary the data were reduced using the ESO MUSE pipeline \citep{Weilbacher_2012} in the ESO Recipe Execution Tool (EsoRex) environment \citep{ESOREX}. The reduction was carried out using the standard procedure, which included creating master bias, flat field and wavelength calibrations for each CCD and for each night of observations, which were then applied to the raw science and standard-star observations as part of the pre-processing steps. The data for each night was then flux calibrated using the standard stars for each night, and the sky background was measured from the science data after masking out \Leda\. and the extended emission from W2246–0526. As part of the sky subtraction process, we also applied the self-calibration step within EsoRex, which applies an autocalibration on the slice-level to smooth the sky background across the entire FOV. Finally, the data were stacked, and the the Zurich Atmosphere Purge code \citep[ZAP,][]{Soto_2016} was used to remove any residual sky contamination from the final datacube.

The white-light image of the galaxy, created from the MUSE datacube, is shown in the left panel of Fig.~\ref{fig:images}, alongside two colour images. The middle image was created using the white-light image centered on H$\alpha$ in red, [\ion{O}{iii}] in green and the stellar continuum in blue, whereas the rightmost image shows the continuum-subtracted H$\alpha$, [\ion{O}{iii}] and [\ion{N}{ii}] emission in green, blue and red, respectively. One can also see that there is a smaller galaxy on the upper edge of the datacube that shows gas emission, particularly in [\ion{O}{iii}], at a similar redshift as \Leda\ . A foreground star also lies on the eastern edge of \Leda\ , and this region was masked out during the analysis to prevent contamination.

\section{Morphology}\label{sec:morphology}
The first step in the analysis of \Leda\. was to consider its morphology and physical properties. As can be seen in Fig.~\ref{fig:images}, \Leda\. is a flocculent spiral galaxy of class Sc, based on a visual inspection of the data used in this study. We used the MUSE datacube to create images of the galaxy in the $g$ and $r$ bands, and modelled the surface brightness profile using \textsc{galfit} \citet{Peng_2002, Peng_2010}. The best fit was found with a two component model representing a bulge and disc. 
%An example of the fit to the $r$-band image is shown in Fig.~\ref{fig:Galfit}. 
It was found that the bulge dominates the light within a radius of $\sim2\arcsec$ ($\sim3.66$~kpc) in all filters, and the galaxy has a mean Bulge to Total (B/T) light ratio of 0.25, indicating that it is disc-dominated.

We used the magnitudes of the bulge and disc from the \textsc{galfit} fits to calculate the total magnitude of the galaxy in each filter, and used the $g$ and $r$ band magnitudes to estimate the stellar mass using 
\begin{equation} 
	\text{log(} M_*\text{  [M}_\odot \text{])}=0.673\times M_g - 1.108 \times M_r + 0.996
	\label{eq:mass}
\end{equation}
\citep{Ebrova_2025}, where $M_g$ and $M_r$ are the absolute magnitudes in the $g$ and $r$ band filters, respectively. This calculation gave a total stellar mass of $4.6 \pm 1.1\times 10^{10}$~M$_\odot$. The uncertainties in this measurement come from both the uncertainties in the equation and from the magnitudes given by \textsc{galfit}. \citet{Haeussler_2007} and \citet{Haeussler_2013} found that the uncertainties derived by \textsc{galfit} are often substantially underestimated since it assumes that any residual flux is due purely to Poisson noise, and does not take into account the effect of a poorly fitted model or other irregular structures, such as spiral arms. \citet{Nedkova_2021} found that for modelling two components, such as the bulge and disc, the uncertaintes from \textsc{galfit} are of the order of $2-2.5$~times too small, so we followed their guide of multiplying the \textsc{galfit} magnitude errors by a factor of 3 in order to be conservative in the mass estimate above. 
This stellar mass is higher than the mass derived by \citet{Fan_2018} of $1.1 \substack{+0.5 \\ -0.2} \times 10^{10}$~M$_\odot$, which could be due to factors such as the methods used to measure the magnitudes \citep[2-component fits with \textsc{galfit} in this paper versus aperture photometry in][] {Fan_2018}, the number of filters used to derive the magnitudes and therefore the mass (2 versus 18), and the methods used to derive the mass (equation~\ref{eq:mass} versus SED fitting).

\section{Stellar and Gas Kinematics}\label{sec:kinematics}

\begin{figure*}
%\sidecaption
\includegraphics[angle=0,width=1\linewidth]{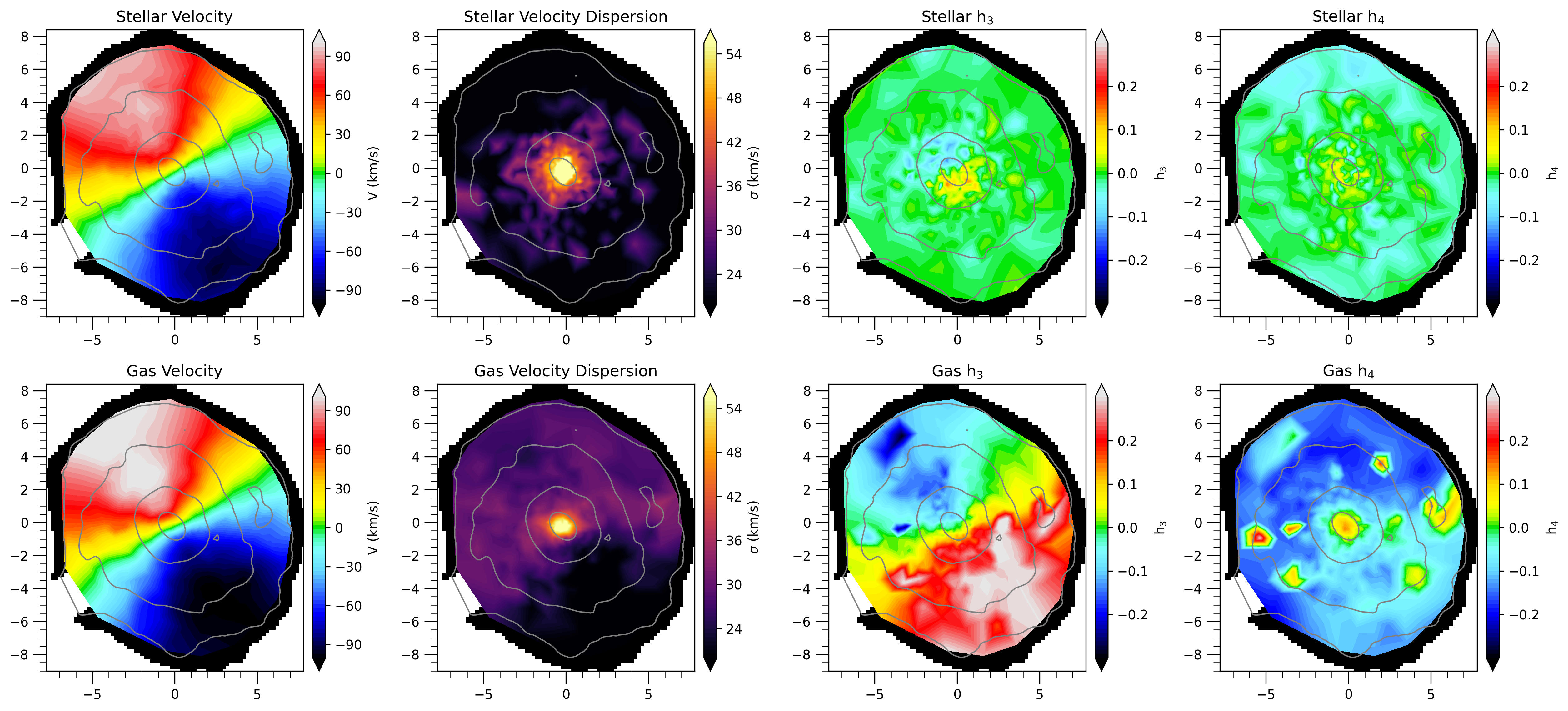}
\caption{Maps of the stellar (top) and gas (bottom) kinematics, showing from left to right the line of sight velocities, velocity dispersions and h$_3$ and h$_4$ coefficients for \Leda\..  The contours represent the flux in the white light image for reference, and the axes labels represent the distance in arcseconds from the centre of the galaxy.
\label{fig:kinematics}}
\end{figure*}

The next step was to determine the connection between \Leda\. and the smaller galaxy to the north by calculating their redshifts using their line of sight velocities. The data was spatially binned using the PowerBin software of \citet{Cappellari_2025}, which applies a 2D adaptive spatial binning to obtain binned spectra with more-or-less the same ``capacity'', in this case signal-to-noise ratio (SNR). For \Leda\. a SNR of 50 in the stellar continuum was selected, and for the fainter companion the data was binned using SNRs of 20, 30 and 40 in order to confirm any detections made at lower SNRs.

Having binned the spectra, the kinematics (line-of-sight velocity, $v$, and velocity dispersion, $\sigma$) for each bin were measured using the penalized Pixel Fitting (pPXF) code of \citet{Cappellari_2023}. For \Leda\. the h$_3$ and h$_4$ coefficients were also measured. The stellar spectra were modelled using template spectra from the E-MILES stellar library \citep{Vazdekis_2016}, which were created using the Padova isochrones and Salpeter IMF and cover a wide range of well-defined ages ($0.6 < \text{Age} < 15.8$ Gyrs) and metallicities ($-1.71 < \text{[M/H]} < +0.22$). Simultaneously, the gas emission lines (H$\gamma$, \ion{He}{ii}$\lambda4687$, H$\beta$, [\ion{O}{iii}$]\lambda5007$, \ion{He}{i}$\lambda5876$, [\ion{O}{i}$]\lambda6300$, H$\alpha$, [\ion{N}{ii}$]\lambda6583$, [\ion{S}{ii}$]\lambda6716$, [\ion{S}{ii}$]\lambda6731$]) were modelled using Gaussian emission line spectra. All the templates for the stellar and gas components were convolved with line-of-sight velocity distributions and velocity dispersions to obtain a model spectrum that best fits each binned galaxy spectrum, and additive polynomials of order 6 were applied to model the flux calibration of the continuum and reduce template mismatch. The fits were run twice, first with the Balmer and forbidden lines constrained to have the same kinematics, and the second time they were modelled as two distinct kinematical components. However, no significant difference was found in their kinematics when they were modelled separately, and so the plots shown in this paper present the results  from the fit using a single gas component that includes all the Balmer and forbidden lines.

The maps for the stellar and gas kinematics for \Leda\.  are shown in Fig.~\ref{fig:kinematics}. These measurements were derived through fits to the spectra over the optical region ($4700 < \lambda < 8100$~\AA, which corresponds to a rest frame wavelength range of $4303 < \lambda < 7409$~\AA), but no significant differences were seen when modeling shorter ranges of the stellar continuum or individual emission lines. It can be seen that both the stars and the gas display a uniform and consistent velocity map, indicating that the galaxy has not undergone any significant mergers or interactions in the last $\sim0.5-1$~billion years \citep{Hung_2016, Davis_2016} that would disrupt the stars and gas and leave a signature in their kinematics.

Both the gas and the stars show a peak in the velocity dispersion at the centre of the galaxy, with the stellar $\sigma$ being slightly higher at the center of the galaxy. One interesting thing to note is the band of slightly higher gas $\sigma$ across the minor axis of the galaxy. This feature is consistent with the ``B distribution of $\sigma$'' that was described by \citet{Pilyugin_2021}. They found that $\sim13\%$ of the MaNGA galaxies in their sample showed this band of higher $\sigma$ across the minor axis of the galaxies. They theorized that this feature could appear in cases where the gas velocity dispersion is asymmetric, i.e. where the radial component of the gas velocity dispersion (i.e. along the line of sight) is greater than the azimuthal (orbiting the galaxy plane) and vertical (perpendicular to the galaxy plane) components. However, they find no evidence that this feature is related to the presence of an AGN. Another study by \citet{Marconcini_2025} also found evidence for the B-distribution in the active galaxy NGC~424, lying perpendicular to the high-velocity  outflow. They theorised that this outflow  may enhance the B distribution of $\sigma$ by injecting energy into the host disc and perturbing the ambient material. However, a full understanding of the origin of this effect is still missing in the literature, and so we report it here as simply another galaxy in which the feature is detected.

The h$_3$ and h$_4$ Gauss-Hermite coefficients are also interesting. These coefficients represent the skewness and kurtosis, respectively, and represent the non-Gaussianity in the shapes of the absorption and emission lines. The stellar h$_3$ map shows a clear anti-correlation with the velocity map in the inner region of the galaxy with some areas of slightly enhanced h$_4$, which together likely represents the presence of two kinematic components- a cold, rapidly rotating disk and a hotter, slower component, such as a bulge. The gas h$_3$ also shows a very strong anti-correlation with the gas velocity, while the h$_4$ coefficient is generally very negative with a few regions of stronger positive values. These maps indicate that the gas is orbiting as a dynamically cold disc, but is more structurally complex than the stellar disc, likely due to turbulence induced along the spiral arms or star-forming regions. These results, particularly the anti-correlation between velocity and h$_3$, are consisent with the predictions of \citet{Naab_2014} for galaxies that built up their mass through in-situ star formation along with gas-rich mergers, and with Class 3 and 4 galaxies (fast rotators with a strong disc component) observed by \citet{vandeSande_2017} in the SAMI survey.

\begin{figure}
%\sidecaption
\includegraphics[angle=0,width=1\linewidth]{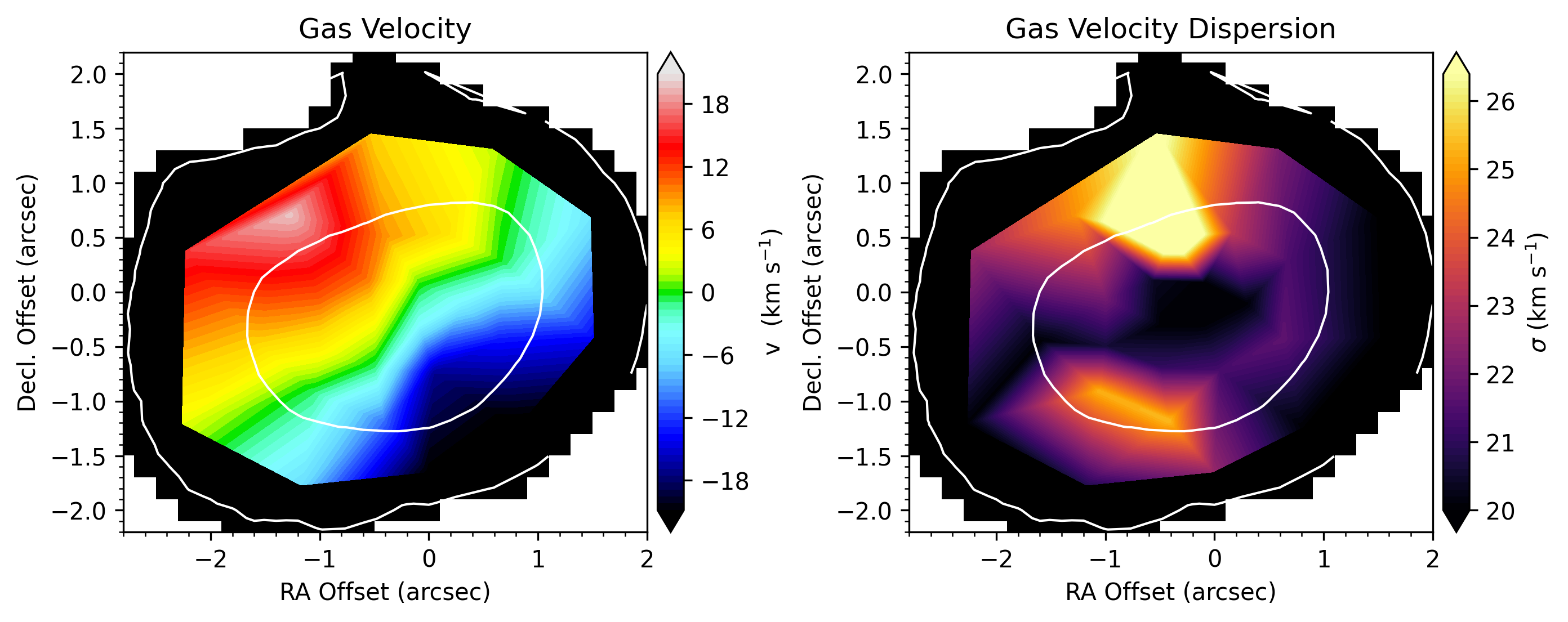}
\caption{Maps showing the stellar line of sight velocities (left) and velocity dispersions (right) for the neighbouring galaxy to \Leda\..  The contours represent the flux in the white light image for reference.
\label{fig:kinematics_sat}}
\end{figure}

The kinematics analysis outlined above was then repeated for the  neighbouring galaxy. While the stellar continuum was too faint to obtain reliable spatially resolved kinematics measurements, it was possible to combine the spectra from the whole galaxy into a single spectrum to derive the mean stellar kinematics. The gas emission, on the other hand, was bright enough to derive kinematics maps for a SNR of 20. These maps are shown in Fig.~\ref{fig:kinematics_sat}. One can see a smooth velocity curve, indicating that this galaxy is rotating along a similar axis as \Leda\.. There are also two peaks in the velocity dispersion plot. While these peaks are only small compared to the velocity dispersion of the rest of the galaxy, they appear close to two regions of strong [\ion{O}{iii}] emission shown in Fig.~\ref{fig:images}, and were also seen when the datacube was binned to different SNR ratios. They are also both offset from the center of the galaxy, which perhaps rules out ionization from an AGN. Thus, it is likely that these regions reflect areas of enhanced star formation \citep{Green_2010,Green_2014,Yu_2019}.

% Having measured the kinematics of both galaxies, we then calculated their distances and physical separation. The mean stellar velocities for Leda~1044720 and the neighbouring galaxy were found to be $26398.0 \pm XX$~km~s$^{-1}$ and $26325.16 \pm XX$~km~s$^{-1}$, respectively, which correspond to distances of $377.11 \pm XX$~Mpc and $376.07 \pm XX$~Mpc, respectively, and redshifts of $0.0922 \pm XX$ and $0.0920 \pm XX$, respectively. The two galaxies have a separation on sky of $\sim19.5\arcsec$, which corresponds to a physical separation on sky of $\sim35.5$~kpc at the distance of Leda~1044720

Having measured the kinematics of both galaxies, we then calculated their redshifts and physical separation. \Leda\. was found to lie at a redshift of $0.09223 \pm 0.00009$ while the satellite has a redshift of $0.09200 \pm 0.00009$, which correspond to distances of $377.1 \pm 0.3$~Mpc and $376.1 \pm 0.3$~Mpc, respectively. The two galaxies have a separation on sky of $\sim19.5\arcsec$, which corresponds to a physical separation on sky of $\sim35.5$~kpc at the distance of \Leda\.. Taking into account their distances from us, we found that they have a physical separation of $1.0 \pm 0.7$~Mpc. However, it should be noted that this distance is an upper limit on their separation assuming that their line-of-sight velocities correspond directly to their distance to us and relative to one another, i.e. assuming zero proper motion between them.

In the literature, galaxies are considered to have a close companion (i.e. to be a pair) or to be interacting when their separation is less than 30~kpc \citep[e.g.][]{Michel_2008, Ellison_2008, Robaina_2009, Patton_2011, Patton_2024}, in some cases up to $100-150$~kpc \citep[e.g.][]{Nikolic_2004, Li_2008, Casteels_2013,Shah_2022,Byrne_2024}. Thus, with a separation of $\sim1$~Mpc, it is unlikely that the two galaxies are interacting or form a gravitationally bound pair. However, it is possible that they reside in the same group or cluster environment, where other members of the group lie outside of the field of view of the data \citep[e.g.][]{Chernin_2000}. Thus, for the rest of this paper we will focus our analysis on \Leda\ .

\section{Stellar Continuum}\label{sec:stellar_cont}
\subsection{Stellar Populations}\label{sec:stellar_pops}

The next step in the analysis was to study the stellar populations across \Leda\. in order to understand the star formation history of the galaxy. This analysis was carried out through a study of the mass- and luminosity-weighted stellar populations. Since the youngest stars are the hottest and brightest stars in the galaxy, they dominate the light even though they only account for a small fraction of the mass. Thus, the luminosity-weighted ages and metallicities give insight into the most recent episode of star formation since this phase created these hot, bright stars. However, the majority of the mass of a galaxy lies in older, fainter and less massive stars, and so a study of the mass-weighted stellar populations gives a better overview as to how the mass of the galaxy built up over time, and is less biased towards more recent episodes of star formation that dominate the light. Together, the luminosity and mass-weighted stellar populations allow us to build up a clearer picture of the star-formation history across the galaxy.

\begin{figure*}
\includegraphics[angle=0,width=0.5\linewidth]{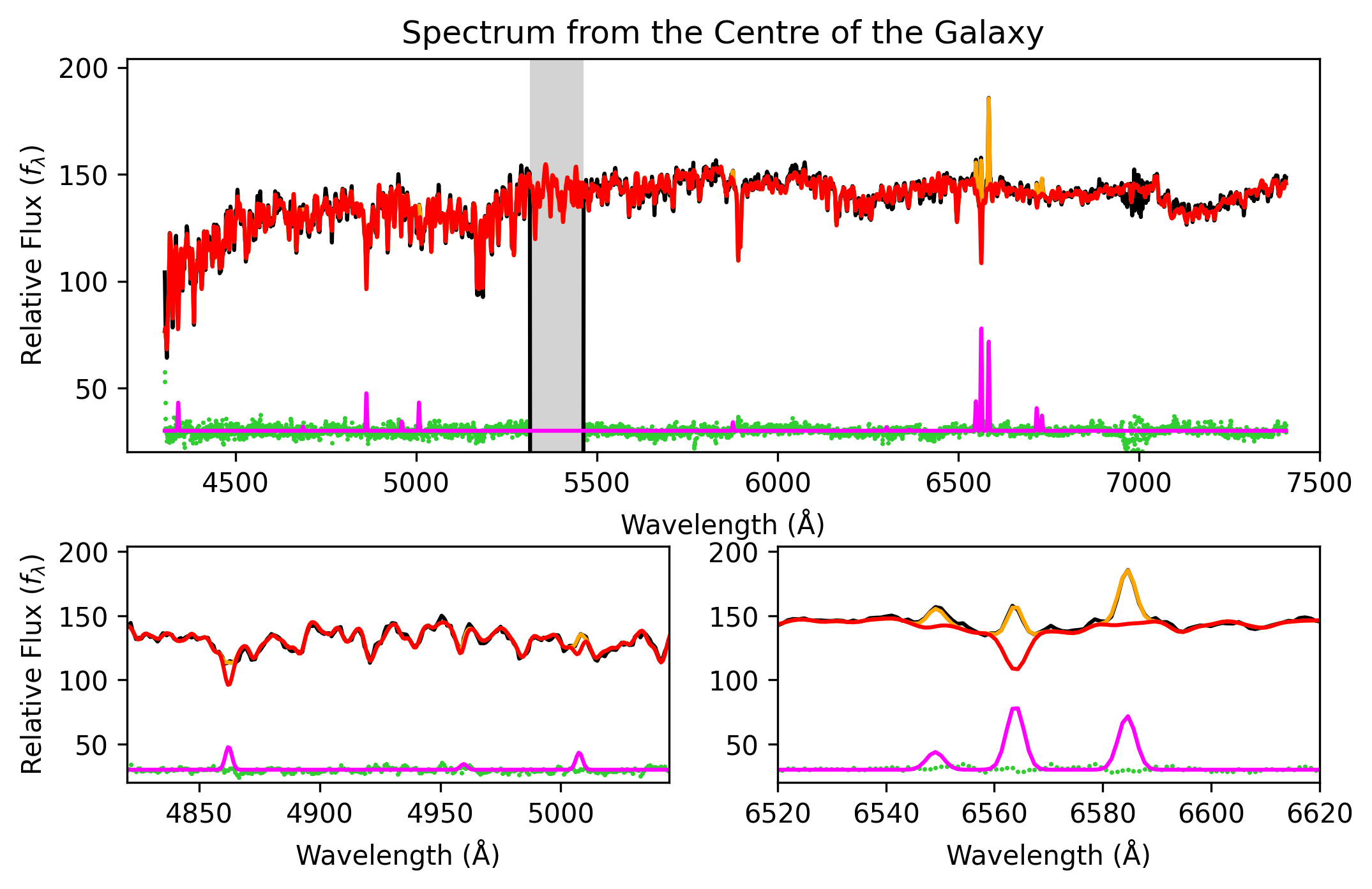}
\includegraphics[angle=0,width=0.5\linewidth]{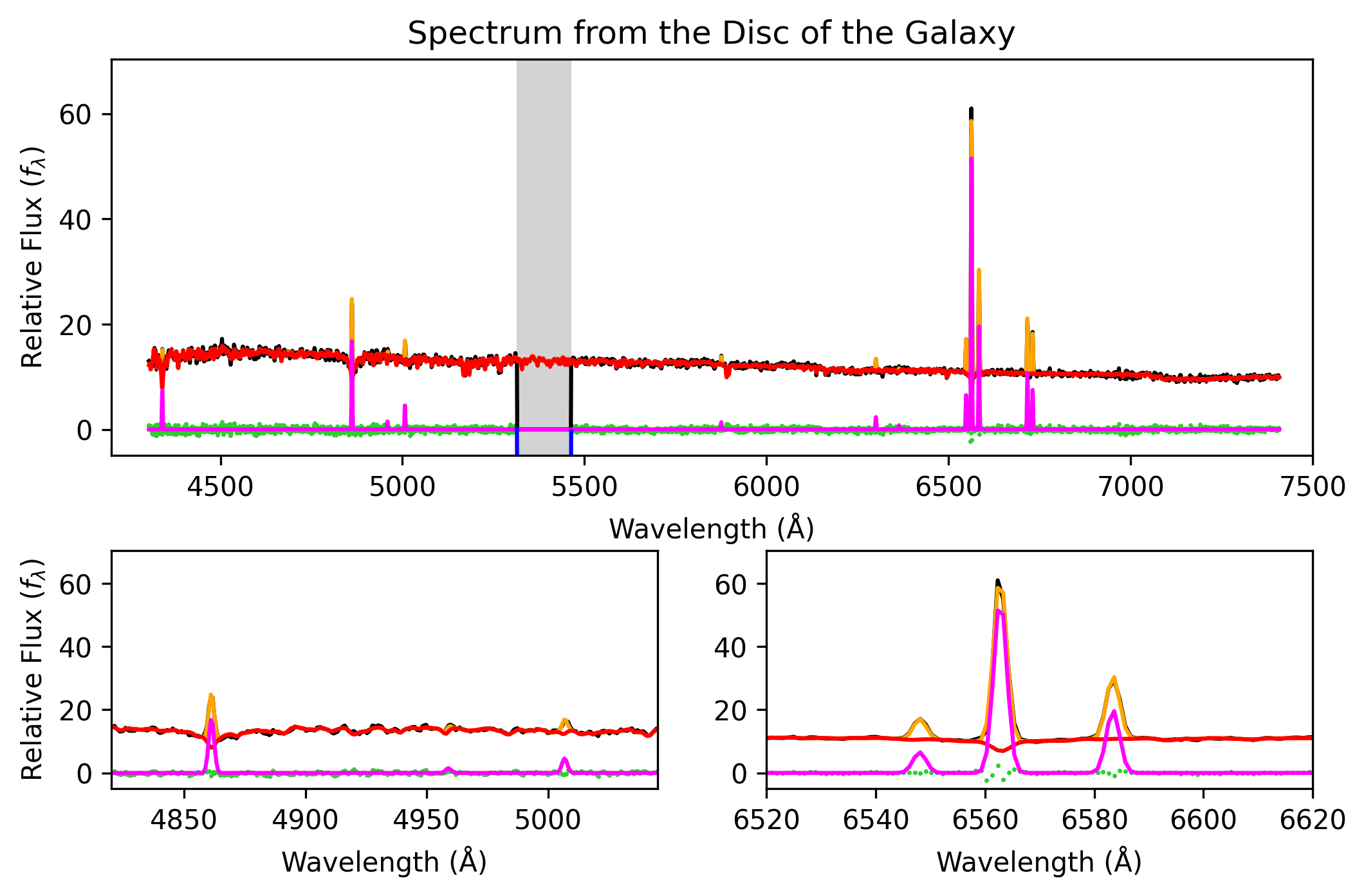}
\caption{Example fits with pPXF to two spectra from the galaxy. On the left is a spectrum taken from close to the center of the galaxy where the amplitude-to-noise ratio for the H$\alpha$
line is weakest, and on the right is a spectrum from the disc. The top plots show the fits over the entire wavelength range, and the plots below show a zoom-in on the H$\beta$ and [\ion{O}{iii}] emission lines on the left and the H$\alpha$ and [\ion{N}{ii}] emission lines on the right. In all plots, the black line represents the binned spectrum from the galaxy, the red and magenta lines are the best fits to the stars and gas, respectively, the orange line shows the combined best fit, and the green points are the residuals. Note that the emission lines and residuals have been offset to higher values to better display the fits.
\label{fig:ppxf_fits}}
\end{figure*}

For this analysis we took the binned spectra from the kinematics analysis and used pPXF to derive estimates of the mass- and luminosity-weighted ages and metallicities. The first step was to correct for the galactic extinction. Since this galaxy has a very high galactic latitude, the galactic foreground redenning is relatively small, with a value of $E(B-V) = 0.0292$ calculated from \citet{Schlafly_2011} using the IRSA Galactic dust reddening tool \footnote{https://irsa.ipac.caltech.edu/applications/DUST/}. We then applied the \citet{Fitzpatrick_1999} extinction curve to calculate the extinction in magnitudes, $A_\lambda$, which was then applied to the binned spectra.

For the fits with pPXF we used the same templates and wavelength range as for the kinematics measurements, described in Section~\ref{sec:kinematics}. The fits were convolved with multiplicative polynomials of order 6, 8 and 10 to account for the shape of the continuum, reducing the sensitivity to dust reddening, and omitting the requirement of a reddening curve. Upon comparing the results derived using each order polynomial, it was found that the ages and metallicities start to converge to consistent values for multiplicative polynomials of orders 8 and 10, whereas those derived using an order of 6 tended to significantly overestimate ages and metallicities, which is consistent with a scenario where the continuum is underfitted  \citep{Cappellari_2017,Cappellari_2023}. Therefore, we selected a multiplicative polynomial of order 8 for the analysis presented in this paper since it provides the optimal fit without underfitting the continuum or risking oversmoothing the spectrum such that any broad absorption features are included in the fit to the continuum. The fits were also carried out using regularization, which works to create a smoothed variation in the weights of the templates with similar ages and metallicities. To determine the level of regularization, we took a spectrum from the center of the galaxy with a high SNR and carried out an unregularized fit to measure the $\chi^2$ value. The noise spectrum was then scaled  until $\chi^2/N_{DOF} = 1$, where $N_{DOF}$ is the number of degrees of freedom in the fit, which corresponds to the number of unmasked pixels in the input spectrum. We then repeated the fit to the spectrum using this scaled noise spectrum and increasing value for the regularization parameter until the $\chi^2$ of the fit increased by $\Delta\chi^2 = \sqrt{2\times N_{DOF}}$. This value represents the limit between a smooth fit that still reflects the star-formation history of the galaxy and one that has been smoothed excessively. It should be noted at this point that this smoothed fit may not  reflect the true star-formation history of that part of the galaxy, which is likely to vary over shorter timescales than the models allow, but instead acts to reduce the age-metallicity degeneracy between spectra, thus allowing a more consistent comparison of systematic trends in their star-formation histories. Having determined the best regularization value for this spectrum, this value was then applied to the fit to the remaining spectra of the galaxy. An example of these fits to two of the binned spectra are presented in Fig.~\ref{fig:ppxf_fits}, showing the fit to a spectrum close to the centre of the galaxy and another one further out in the disc. 
%Since only a single estimate of age and metallicity was required from these fits, they were carried out without regularization as they are computationally less expensive and have been found to derive consistent values for average age and metallicity \citep[e.g.][]{Boardman_2017, Carrillo_2020, Lu_2023}.

\begin{figure*}
\includegraphics[angle=0,width=1\linewidth]{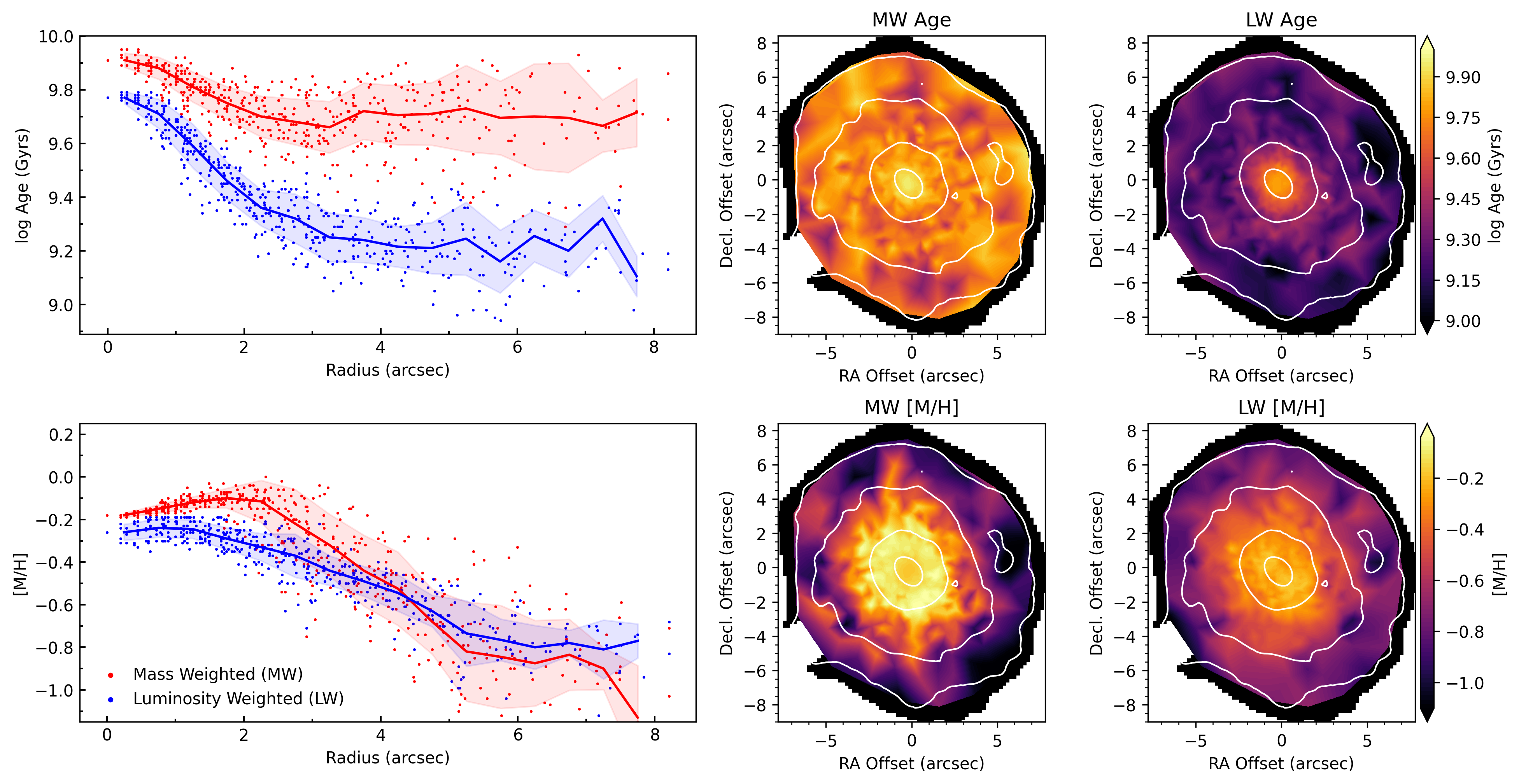}
\caption{Overview of the mass- and luminosity-weighted ages (top) and metallicities (bottom). On the left are radial plots for these properties, with the red and blue points representing the measurements from each binned spectrum for the the mass- and luminosity-weighted stellar populations, respectively. The solid lines represent the rolling average for each data set, and the shaded areas show the 1~$\sigma$ variation in this average. On the right are the mass and luminosity-weighted stellar population maps, where the contours reflect the variation in the white light image created from the datacube. 
\label{fig:radial_plots_stars}}
\end{figure*}

The weights for each template spectrum in each fit were then used to calculate the mean luminosity and mass-weighted ages and metallicities for each binned spectrum using 
\begin{equation} 
	\text{log(Age)}=\frac{\sum \omega_{i} \text{log(Age$_{\text{template},i}$)}}{\sum \omega_{i}}
	\label{eq:age}
\end{equation}
and 
\begin{equation} 
	\text{[M/H]}=\frac{\sum \omega_{i} \text{[M/H]}_{\text{template},i}}{\sum \omega_{i},}
	\label{eq:met}
\end{equation}
respectively, where $\omega_{i}$ represents the weight of the $i^{th}$ template (i.e. the value by which the $i^{th}$  stellar template is multiplied to best fit the galaxy spectrum), and [M/H]$_{\text{template},i}$ and Age$_{\text{template},i}$ are the metallicity and age of the $i^{th}$ template respectively. The results are shown in Fig.~\ref{fig:radial_plots_stars}, with the ages on the top row and metallicities on the bottom row. On the left are the radial plots, with the mass- and luminosity-weighted measurements in red and blue, respectively, and on the right are the radial maps for these properties.

Considering first the stellar ages, one can see that the mass-weighted ages are relatively flat across the galaxy, with slightly older stars at the centre. This result is consistent with a galaxy that has undergone relatively constant star formation across the whole disc over most of its life, with an older bulge at the center that formed early and has undergon little star formation since then. Due to the relatively old ages and small radial differences at radii larger than $\sim 2\arcsec$, where the disc dominates the light, it is not possible to say whether the disc of the galaxy has formed inside-out or outside-in. On the other hand, the luminosity-weighted ages show a clear negative gradient in the inner region, within a radius of $\sim2-3\arcsec$ where the ages drop from around $6-7$ billion years at the core to 1.5~billion years, and then a flatter gradient in the rest of the disc. This result is consistent with a galaxy that has ongoing star formation throughout the disc, and where the star formation in the inner region has been truncated somehow, leading to older stars there. From the light profile fitting described in Section~\ref{sec:morphology}, we found that the bulge dominates the light at the centre of the galaxy, and the disc starts to dominate the light at a radius of $\sim2\arcsec$. This radius corresponds to approximately to the second contour from the centre of the galaxy shown in Fig.~\ref{fig:kinematics} and in subsequent figures. Therefore, it is possible that this trend in the luminosity-weighted ages reflects the dominance of the older bulge stellar populations at the very centre of the galaxy, and with the increasing proportion of disc light as the radius increases. 

However, despite the contribution of the bulge, the disc itself could also show a slight negative age gradient. Many studies have found negative age gradients in spiral discs. For example,  \citet{Goddard_2017} and \citet{Pessa_2023} found negative luminosity-weighted age gradients in a large sample of MaNGA and PHANGS-MUSE galaxies, respectively. \citet{Pessa_2023} also found negative mass-weighted age gradients in their galaxies, which they attributed to inside-out formation, where the inner regions of the galaxy formed earlier than the outskirts. 

The flat mass- and luminosity-weighted age gradients across the disc may indicate one of several growth scenarios. For example,  it may reflect that the galaxy did not build up its mass through a slow inside-out growth mechanism since this process would typically produce a strong negative age gradient. Or, if such a process did occur, it happened long enough ago that the age gradient has been smoothed out due to continuous star formation and is now difficult to detect. \citet{SanchezB_2014} also found flatter mass-weighted age gradients across the discs of their sample of CALIFA galaxies, with typical ages around $\sim10$~Gyrs, suggesting that another factor could be radial mixing over the lifetime of the galaxy.

Moving onto the stellar metallicity, it can be seen that both the mass- and luminosity-weighted metallicity curves show a negative gradient. Interestingly, the mass-weighted metallicities increase slightly from the centre out to a radius of $\sim2\arcsec$ before decreasing again towards the edges of the galaxy. This drop at the very centre aligns with the region of older luminosity-weighted stellar populations, and is consistent with the scenario where in that region the star formation has been truncated, reducing the chemical enrichment there relative to other parts of the disc that have ongoing star formation.

The decrease in stellar metallicity from a radius of $\sim2\arcsec$ towards the outskirts of the galaxy could reflect longer star formation in the inner disc than the outer disc, which would lead to higher chemical enrichment of the ISM through supernovae or stellar winds \citep[e.g.][]{Lara_2022}, or perhaps accretion of pristine material into the outskirts of the disc \citep[e.g.][]{Aragon_2013, Aragon_2019,Egorova_2025}, thus reducing the metallicity there. In the outermost regions of the galaxy the luminosity-weighted metallicities can be seen to be slightly higher in general than the mass-weighted metallicities. This region coincides with the younger luminosity weighted ages, and is consistent with more recent or more active star formation in that region producing younger stars and higher chemical enrichment. Negative metallicity gradients, and also negative age gradients,  are common in spiral galaxies \citep[e.g.][]{Lian_2018, Parikh_2021, Pessa_2023}, with steeper gradients seen in more massive and late type galaxies similar to \Leda\. \citep{SanchezB_2014, Goddard_2017}, and are often attributed to inside-out formation or growth of the galaxy disc \citep{Tissera_2016, Breda_2020}.

% \subsection{Stellar Mass}\label{sec:stellar_pops}
% The full spectral fitting method described above can also be used to calculate the total stellar mass of the galaxy. First we created the mean spectrum of the galaxy and applied the fit with pPXF to derive the luminosity-weighted stellar populations. Then we used the mass-to-light ratios of each template included in the fit along with the magnitudes derived from the fits with Galfit in Section~\ref{sec:morphology} to derive the total stellar mass. For the $r$-band, we found a stellar mass of $4.75 \pm 0.05\times 10^{10} M_\odot$, which is consistent with the mass derived using only the $g$ and $r$-band magnitudes in Section~\ref{sec:morphology}.

\section{Gas Properties }\label{sec:gas_properties}
The stellar populations, particularly the stellar metallicity, tell us about the long term state of the ISM enrichment over the last $1-10$~billion years, in particular about the conditions within the galaxy at the time that the stars were created \citep[e.g.][]{Gibson_1997}. However, the ISM is continuously enriched through stellar evolution processes such as Type II supernova explosions, stellar winds from massive and AGB stars, and ongoing star formation. As a result, to understand the current chemical enrichment of the ISM, one must instead examine the properties of the ionized gas, which trace enrichment processes operating on relatively short timescales of $10-100$~million years \citep{Mannucci_2005, Scannapieco_2005, Sullivan_2006, Matteucci_2009, Brandt_2010, Maoz_2011, Maoz_2012}.

\subsection{BPT diagrams}\label{sec:BPT}

\begin{figure*}
\sidecaption
\includegraphics[width=12cm]{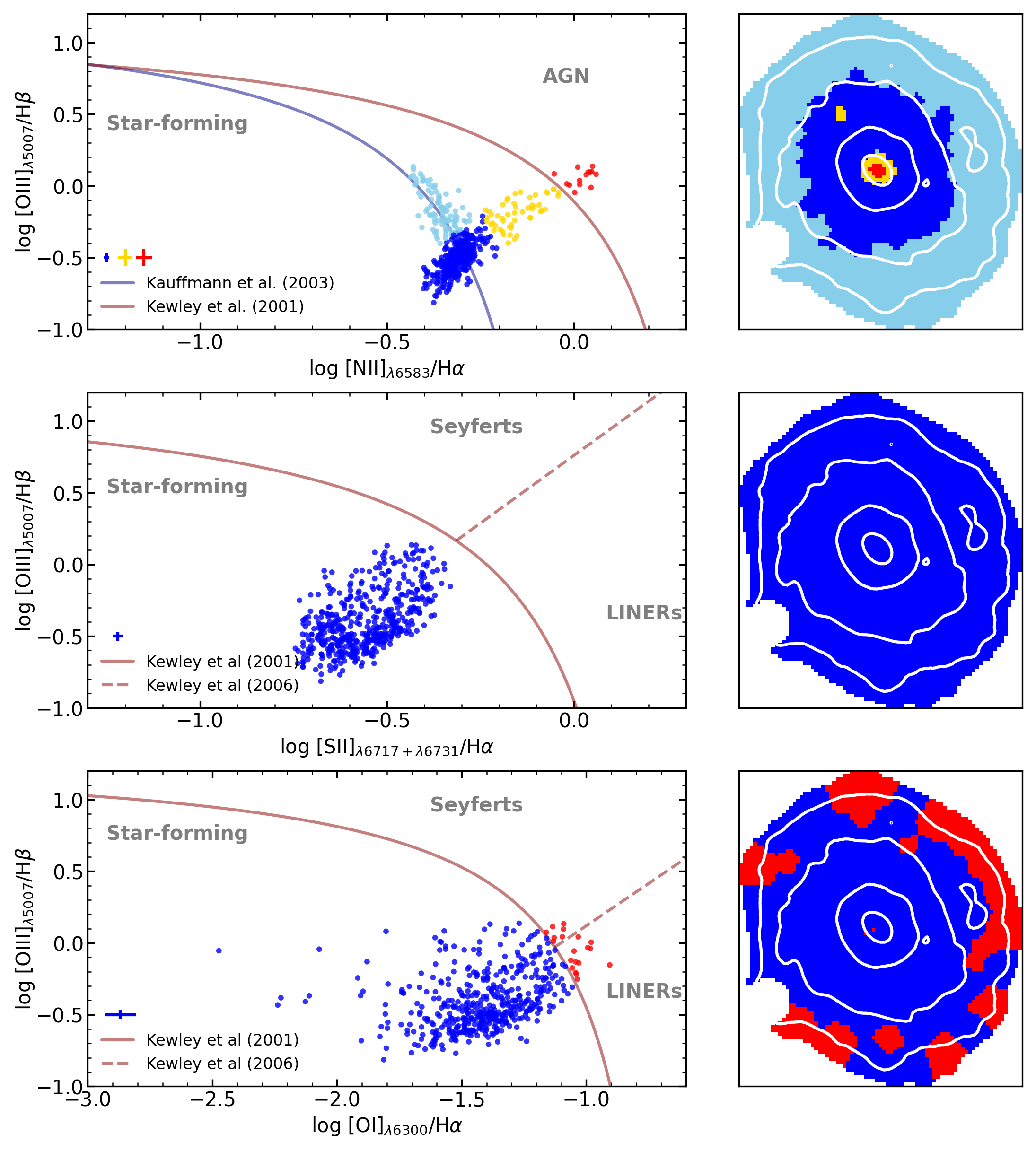}
\caption{ The BPT diagram for \Leda\, at the top along with the [\ion{S}{ii}] (middle) and [\ion{O}{i}] diagnostic diagrams on the left. The brown curves represent the maximum starburst line of \citet{Kewley_2001}, the dashed brown lines represent the separation between Seyferts and LINERs from \citet{Kewley_2006}, and the blue curve in the top plot represents the pure star-forming line of \citet{Kauffmann_2003}.
%The dashed line in the bottom plot represents the separation between Seyferts and LINERs, where shock ionized regions often overlap with the LINER region in this plot as well. 
The error bar above the legend represents the median uncertainties in the line ratios. The uncertainties in the BPT diagram at the top have been separated according to the binned spectra that lie in the star-forming, composite and AGN regions of the diagram, and use the same colour scheme.
The  maps on the right demonstrate the distribution of the bins in each region of the diagnostic diagrams over the galaxy. 
% The black points in the radial plots represent the measurements from the neighbouring galaxy.
}
\label{fig:BPT}
\end{figure*}

The first step in analysing the gas properties was to identify the main source of ionization across the galaxy, such as star formation or AGN, using variations of the BPT diagram of \citet{Baldwin_1981}. For each binned spectrum, the emission lines were modelled alongside the stellar continuum with pPXF, which corrected the emission line fluxes for any absorption lines present at the same wavelengths.  The emission line ratios were then calculated and plotted onto the BPT diagram shown in the top-left panel of Fig.~\ref{fig:BPT}. The mean uncertainty for the spectra in each region of the BPT diagram (star-forming, composite and AGN) are shown above the legend, having been calculated from the fits with \textsc{ppxf}.

The top plot in Fig.~\ref{fig:BPT} shows the BPT diagram for [\ion{N}{ii}]/H$\alpha$ and [\ion{O}{iii}]/H$\beta$, which is sensitive to metallicity and ionization state. The data here displays the characteristic ``Y'' shape seen in galaxies with both star formation and an AGN present. To better understand this BPT diagram, the BPT map of the galaxy is shown on the right of Fig.~\ref{fig:BPT}, where the colours represent the bins in different parts of the BPT diagram on the left.  The majority of the disc lies on the star-forming branch, which has been separated into the upper and lower parts in the BPT diagram (coloured light and dark blue, respectively). The inner part of the disc lies on the lower part of this branch in the BPT diagram while the outer part of the disc lies towards the top of the star-forming branch. The shape of the star-forming branch in the BPT diagram is driven by many factors, such as the gas-phase metallicity, the stellar mass, star-formation rate, the hardness of the ionizing radiation, the ionization parameter, and the relative abundances of N/O versus O/H, to name but a few. These effects don't scale linearly with each other, which produces the curved shape for the star forming branch, such as can be seen in Fig.~\ref{fig:BPT}. However, a study by \citet{Curti_2022} found that, to a first order, the shape of the star-forming branch of the BPT diagram primarily traces the gas-phase metallicity, from higher metallicities at the bottom of the branch to  lower metallicities and more younger O-class stars towards the upper part of the branch. Thus, in this galaxy, the BPT diagram is reflecting higher gas phase metallicities in the inner disc and lower metallicities in the outer disc. This trend is consistent with the stellar metallicity trend seen in Fig.~\ref{fig:radial_plots_stars}, and with the distribution of H$\alpha$ and [\ion{O}{iii}] in Fig.~\ref{fig:images}, where the [\ion{O}{iii}] is strongest in the outer part of the disc. [\ion{O}{iii}] is a high ionization line, and its distribution can suggest the presence of hot stars in the inner regions of the HII regions since these young massive stars tend to form in regions with lower metallicity gas.

The central region of the galaxy lies to the right of the star-forming branch and towards the LINER (red) and composite (yellow) regions of the BPT diagram. Normally, this trend would indicate the presence of an AGN at the centre of the galaxy. However, when comparing to the right panel of Fig.~\ref{fig:images} one can see that this central region of the galaxy contains very strong [\ion{N}{ii}] emission but very weak H$\alpha$ emission, which might be enhancing the [\ion{N}{ii}]/H$\alpha$ line ratios in that part of the galaxy. Similarly, it can be seen that the uncertainties for the data points in this region are larger than in the star-forming branch, potentially also indicating that at least one of the line strengths is getting weaker and that the uncertainty is proportionally larger.

Two variations of the BPT diagram are regularly used to help further refine the dominant ionizing sources across galaxies. These diagrams use the [\ion{S}{ii}]/H$\alpha$ and [\ion{O}{i}]/H$\alpha$ line ratios against [\ion{O}{iii}]/H$\beta$ \citep{Veilleux_1987, Kewley_2001, Kewley_2006}, hereafter referred to as the [\ion{S}{ii}] and [\ion{O}{i}] diagnostic diagrams. The [\ion{S}{ii}] diagnostic diagram is used to better distinguish between Seyferts and LINERS, but in the diagram shown in the middle panel of Fig.~\ref{fig:BPT} it can be seen that all the line ratios from across the galaxy lie within the star-forming region of this diagram. The [\ion{O}{i}] diagnostic diagram, on the other hand, is useful for identifying ionization from shocks since it is more sensitive to hard ionizing radiation and partially ionized zones. Typically, if regions of a galaxy have been ionized by shocks, they overlap with the LINER region in the [\ion{O}{i}] diagnostic diagram. However, it can be seen in Fig.~\ref{fig:BPT} that the central region of this galaxy again lies in the star-forming region of the diagnostic diagram, indicating that if shocks are present there, their effect is negligible.

 In general, although the classical BPT diagram is widely used, it has important limitations. The ``composite'' region between star-forming and AGN-dominated zones \citep{Kauffmann_2003, Kewley_2001} can include low-luminosity or metal-poor AGNs, supernova remnants, and even purely star-forming regions \citep[e.g.][]{Agostino_2019, CidFernandes_2021}. Recent studies have also found that galaxies at high redshift, such as the Little Red Dots observed with the JWST, appear to lie in the composite and star-forming regions, likely due to their lower metallicity \citep[e.g.][]{Nakajima_2022, Kocevski_2023, Ubler_2023, Harikane_2023, Maiolino_2024}. And finally, evolved ionizing sources, such as post-AGB stars and hot low-mass evolved stars (HOLMES), can mimic AGN-like line ratios despite weaker emission, and shocks — both fast and slow — can also produce similar ratios depending on gas conditions and shock velocity \citep[e.g.][]{Binette_1994, Dopita_1996, Flores_2011, LopezC_2020}. These ambiguities make interpreting BPT diagrams challenging.

%Thus, from these three diagnostic diagrams alone, it is unclear what the dominant source of ionization is at the centre of \Leda\ ., and so we must consider other diagnostic diagrams to better understand this region.

% The top plot in this figure shows the relation between [\ion{S}{ii}]/H$\alpha$ and [\ion{O}{iii}]/H$\beta$ from \citet{Kewley_2006}, and is a good indicator of whether shocks are present. In this case,  all the points in this plot lie within the star-forming region, indicating that if shock ionization is present, it is at a negligible level. 

\subsection{WHAN and WHaD diagrams}\label{sec:WHAN_WHAD}
In response to the need for a better identification of ionizing sources, we use hybrid diagnostic diagrams that incorporate not only flux intensities but also additional parameters, such as the estimated equivalent width \citep{CidFernandes_2011} and velocity dispersion \citep{Sanchez_2024}. \citet{Stasinksa_2008} found that the LINER region of the BPT diagram also contains galaxies that have stopped forming stars, and which are instead ionized by the hot low-mass evolved stars (HOLMES), such as hot post-asymptotic giant branch stars and white dwarfs, contained in them. In these regions, the emission line strengths would be weaker than those produced by star formation or AGN, but the line ratios would still cover a wide range of values, overlapping with those typically seen in star-forming regions or AGN. They called these galaxies ``retired galaxies'' to differentiate them from passive galaxies, which contain no or very low/undetectable levels of emission. Later studies with IFU data found that these line ratios are not only seen in retired galaxies, but also retired regions within galaxies \citep[e.g.][]{Singh_2013, Belfiore_2017, Lacerda_2018}. However, with the BPT diagram on its own, it is difficult to identify these galaxies or regions.

In order to better distinguish between retired galaxies or regions in galaxies and LINERS, \citet{CidFernandes_2011} proposed the $W_{\rm H\alpha}$ versus [\ion{N}{ii}]/H$\alpha$ (WHAN) diagram. The WHAN diagram uses the equivalent width of H$\alpha$, W$_{H\alpha}$, to separate these two regimes since star forming galaxies and AGN would be expected to have stronger  H$\alpha$ emission lines than retired galaxies. The first step towards plotting the WHAN diagram was to correct the H$\alpha$ flux for dust extinction before calculating the equivalent width. For this step we used the Balmer decrement to measure the level of dust attenuation across \Leda\ . The theoretical value of the Balmer decrement is 2.86 for case-B recombination \citep{Osterbrock_1989, Calzetti_2000} for a fixed electron temperature $T_e = 10,000$~K and density $n_e = 100~$cm$^{-3}$ \citep{Hummer_1987}. Therefore, any deviation from this theoretical value can be considered an effect of dust extinction \citep{Groves_2012}.

\begin{figure}
\includegraphics[angle=0,width=1\linewidth]{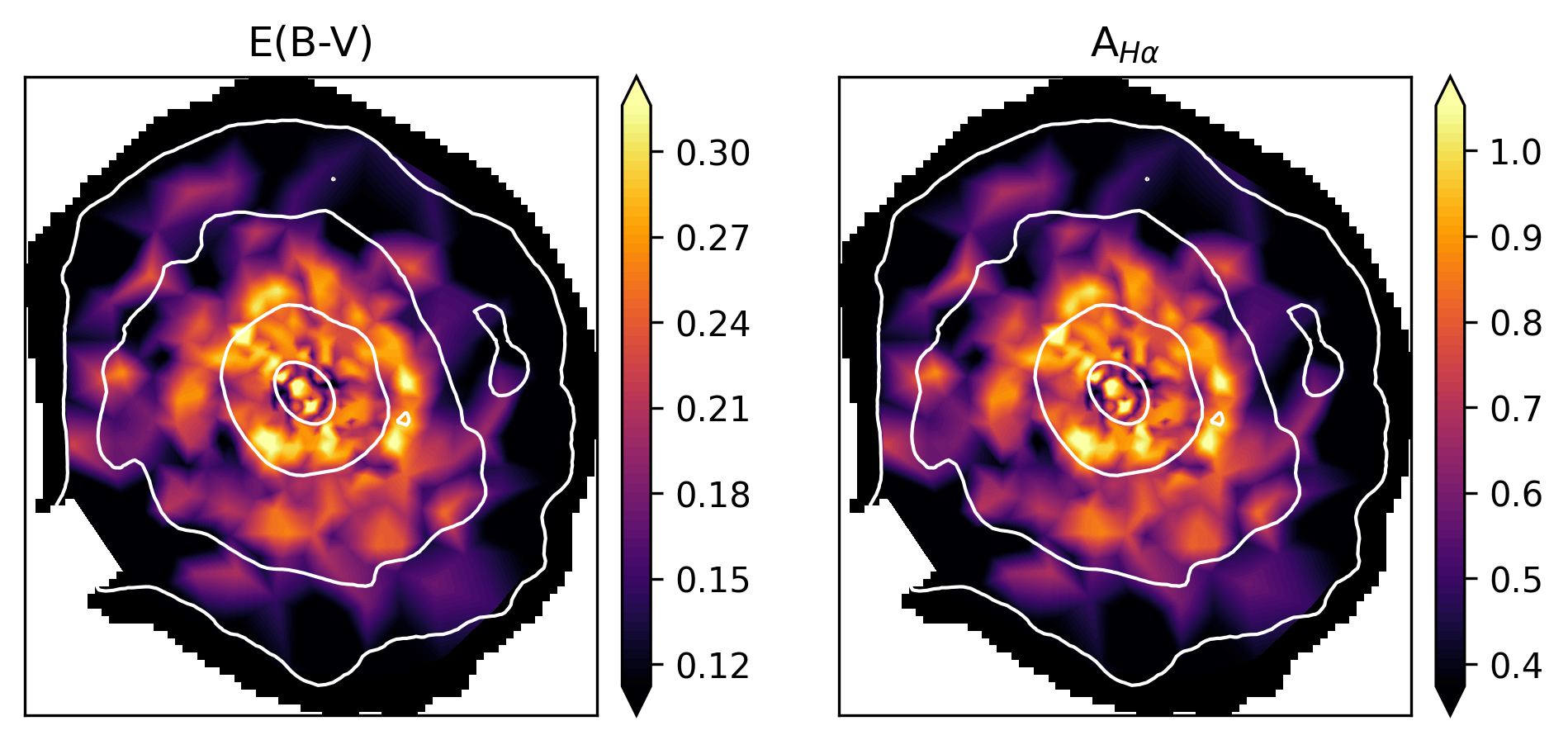}
\caption{Reddening (left) and extinction (right) maps of \Leda\ .  
\label{fig:extinction}}
\end{figure}

We first calculated the foreground dust reddening along the line of sight  using
\begin{equation}
E(B-V) = \frac{2.5}{k(H\beta) - k(H\alpha)} \times\text{log}\Bigg(\frac{F(H\alpha)_o/F(H\beta)_o}{2.86}\Bigg)
\end{equation}
\citep{Dominguez_2013}, where the extinction coefficients $k(H\beta)$ and $k(H\alpha)$ come from the reddening curve described in \citet{Cardelli_1989}, and $F(H\alpha)_{o}$  and $F(H\beta)_{o}$ are the observed H$\alpha$ and H$\beta$ fluxes. We then calculated the extinction at H$\alpha$, $A_{\rm \alpha}$, using 
\begin{equation}
A_{H\alpha} = k(H\alpha) \times E(B-V)
\end{equation}
\citep{Calzetti_2000}. The maps for the dust reddening and extinction are given in Fig.~\ref{fig:extinction}, where it can be seen that the galaxy contains a moderate amount of dust which is mainly concentrated mainly in the inner region of the disc. 

\begin{figure*}
\sidecaption
\includegraphics[width=12cm]{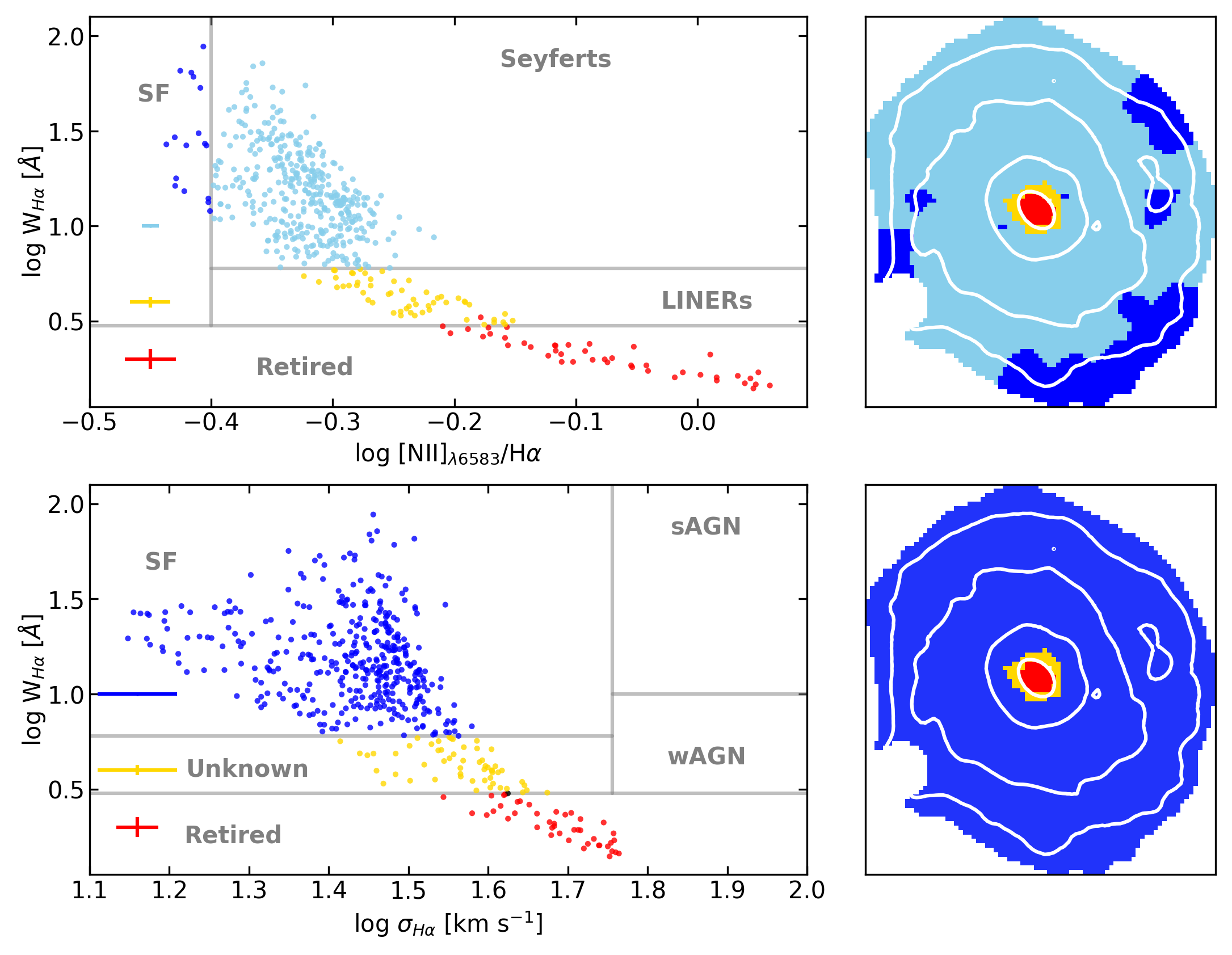}
\caption{The WHAN (top) and WHaD (bottom) diagrams for \Leda\~ on the left and the corresponding maps on the right. The gray lines demonstrate the regions of each diagram that are dominated by different ionization sources, such as star formation (SF), Seyferts (sAGN), LINERS (wAGN) and HOLMES in retired regions of the galaxy. The error bars on the left show the mean uncertainties for the points in each region of the diagram, where the points in the star forming and Seyferts regions of the WHAN diagram have been combined to give a single error. The colours in the plots and the maps reflect which part of the WHAN and WHaD diagram those bins lie in to better show their spatial distribution across the galaxy. }
\label{fig:WHAD_WHAN}
\end{figure*}

Having calculated the dust extinction, we could then correct the observed H$\alpha$ flux and derive the intrinsic H$\alpha$ flux, $F(H\alpha)_{i}$, using
 \begin{equation}
F(H\alpha)_{i} = F(H\alpha)_{o} \times 10^{0.4~A_{H\alpha}}.
\end{equation}
Finally, having corrected the H$\alpha$ flux for dust extinction, we derived the $W_{\rm H\alpha}$ and plotted the WHAN diagram shown in the top panel of Fig.~\ref{fig:WHAD_WHAN}. Since weak emission lines can be hard to model accurately, especially if their amplitude approaches the level of noise in the spectrum, their fluxes and consequently the $W_{\rm H\alpha}$ can be affected. For example, the fluxes of weak emission lines in low SNR spectra could be underestimated, leading to a misclassification of that region as retired. In order to prevent this situation, the amplitude-to-noise ratio (ANR) for the H$\alpha$ line was measured in the fit to each binned spectrum in order to identify potential cases where the H$\alpha$ emission line could be hidden in the noise. In general we found very high ANRs for all the binned spectra, with the minimum ANR being $\sim21$, and the fit to this spectrum is shown in the left panel of Fig.~\ref{fig:ppxf_fits}. As a result, we did not need to mask any points in the WHAN diagram due to unreliable measurements of the emission line flux.

The colours of the data points in the top panel of Fig.~\ref{fig:WHAD_WHAN} correspond to which region of the WHAN diagram they lie in- star forming (SF) in dark blue, Seyferts in light blue, LINERS in yellow and retired in orange-, and the map on the right of the figure shows the distribution of these regions across the galaxy. One can see that the centre of the galaxy has relatively low $W_{\rm H\alpha}$ ($<3 \AA$), putting this region in the retired region instead of the LINER region. This region coincides with the area marked as a LINER and composite region from the BPT diagram in Fig.~\ref{fig:BPT} based on the [\ion{N}{ii}]/H$\alpha$ ratio, and the presence of post-AGB stars is consistent with the old luminosity-weighted stellar populations seen in this region in Fig.~\ref{fig:radial_plots_stars}.

However, this diagram puts the majority of the disc into the LINERS and Seyferts region of the WHAN diagram, which is unlikely to be real. The vertical line separating the star-forming and active galaxies at log~[\ion{N}{ii}]/H$\alpha=-0.4$ corresponds to the optimal transposition of the \citet{Stasinksa_2008} BPT-based SF/AGN division, which is designed to differentiate sources where star formation provides all ionizing photons from those where a harder ionizing spectrum is required. It can be seen in the bottom panel of Fig~\ref{fig:BPT} that the star-forming branch spans both the star-forming and composite regions of the BPT diagram as well. The horizontal position of the star-forming branch can be affected by several factors. For example, galaxies with higher metallicities or higher N/O ratios will typically display higher [\ion{N}{ii}]/H$\alpha$ ratios, pushing them into the composite region on the BPT diagram \citep[e.g.][]{Perez_2009, Belfiore_2017, Curti_2022}. This scenario is likely to be the case for \Leda\. in the BPT and WHAN diagrams.

More recently, \citet{Sanchez_2024} proposed WHaD diagram of $W_{\rm H\alpha}$versus the velocity dispersion of H$\alpha$, $\sigma_{\text{H}\alpha}$, to better differentiate between the different ionizing sources present in a galaxy. In this case, they used a single line, H$\alpha$ for the analysis, making this diagram useful in cases where the other key emission lines are not covered or where the SNR is too low in those lines. The WHaD diagram for \Leda\. is given in the bottom of Fig.~\ref{fig:WHAD_WHAN} along with the spatial distribution of the points. It can be seen that, again, the central region of the galaxy lies in the retired section of the diagram, and the majority of the disc lies in the star-forming part of the diagram. 

Based on these results, it appears that the BPT diagram over-interpreted the hardness of the ionizing radiation field in \Leda\ . While the [\ion{N}{ii}]/H$\alpha$ BPT diagram suggested that the centre of this galaxy contains a LINER, through the analysis of the WHAN and WHaD diagrams it has become clear that in that region of the galaxy the emission is weak and consistent with an old stellar population, rather than with an AGN. Our results are consistent with the category of central  low-ionization emission-line region galaxies  (cLIER galaxies) proposed by \citet{Belfiore_2016}, who defined them as ``galaxies where LIER emission is resolved but spatially located in the central regions, while ionization from star formation dominates at larger galactocentric distances."

\subsection{Gas Metallicity}\label{sec:gas_met}
\begin{figure*}
% \sidecaption
% \includegraphics[width=12cm]{figures/radial_plots_and_maps_gas.png}
\includegraphics[angle=0,width=0.9\linewidth]{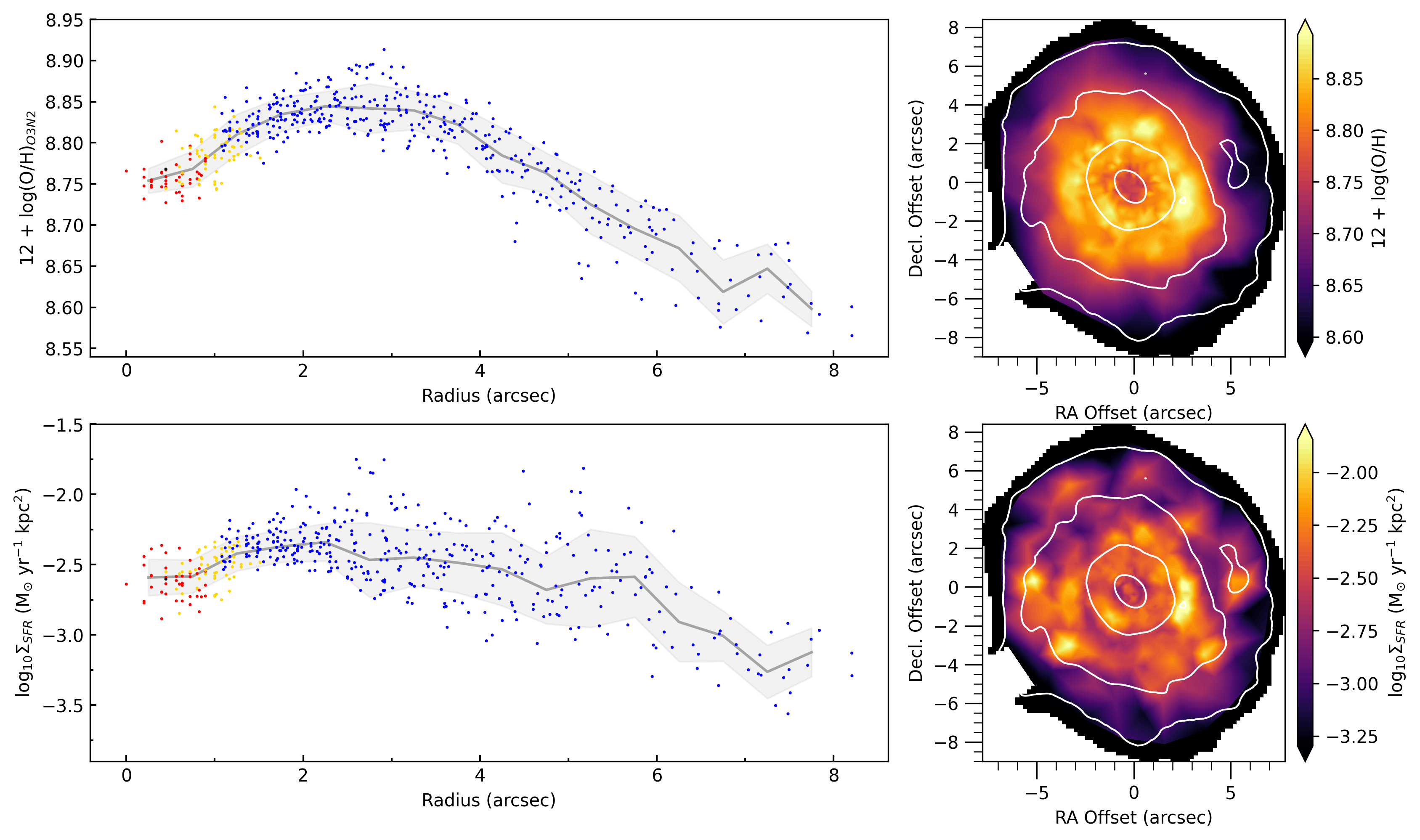}
\caption{The gas metallicity (top) and the SFR density (bottom), plotted as a function of distance from the centre of the galaxy on the left and as a map on the right. In the radial plots, the grey curve shows the rolling average with the shaded area representing the 1~$\sigma$ distribution. The colours of the points in the radial plots indicate the dominant source of ionization in that bin according to the WHaD diagram in Fig.~\ref{fig:WHAD_WHAN}.  
\label{fig:radial_plots_gas}}
\end{figure*}

The shape of the star-forming branch in the BPT diagram is defined by various factors, such as the star formation rate and gas metallicity. In order to study more deeply the relationship between gas metallicity and the position along the star-forming branch on the BPT diagram, we calculated the gas metallicity using the O3N2 indicator:
\begin{equation}
\text{O3N2} = \text{log} \Bigg(\frac{[\text{NII}]6584/\text{H}\alpha}{[\text{OIII}]5007/\text{H}\beta} \Bigg).
\end{equation}
We then calculated the gas phase metallicity using
\begin{equation}
12+\text{log(O/H)}=8.73 - 0.32\times\text{O3N2}
\end{equation}
from \citet{Pettini_2004}. This equation is only valid for the range $-1<\text{O3N2}<1.9$, which is found to be true for all spectra extracted from this galaxy. The radial trend and the map for 12 + log(O/H) is given in the top panel of Fig.~\ref{fig:radial_plots_gas}, alongside the map of the gas metallicity. As a guide, the data points in the radial plots have been colour coded according to the region they lie in the WHaD diagram in Fig.~\ref{fig:WHAD_WHAN}. One can see that as you move outwards from the centre of the galaxy, the gas metallicity increases to a peak at  a radius of $\sim2-3\arcsec$ before decreasing towards the outskirts of the galaxy. In order to confirm that this effect is real, the gas-phase metallicity was calculated again from the O3N2 ratio using the equation of \citet{Marino_2013}, and the same peak at this radius was observed.

The negative trend across the outer disc has been seen in many star-forming galaxies \citep[e.g.][]{Zaritsky_1994, Moustakas_2010, Rupke_2010, Sanchez_2014, Lian_2018, Franchetto_2021, Vale_2025} and in simulations \citep[e.g.][]{Tapia_2025}, and is often attributed to a radial gas inflow, where the gas that is migrating inwards is being continuously enriched by ongoing star formation as it moves through the galaxy \citep{Wang_2022}. This negative metallicity gradient across the disc is consistent with the star-forming branch in the BPT diagram in Fig.~\ref{fig:BPT}, in which the inner and outer parts of the dis sit on the lower and upper regions of this branch, respectively, which corresponds to higher and lower metallicities, respectively.

The inner region of the galaxy ($<3\arcsec$), on the other hand, shows a positive gradient, where the gas metallicity decreases towards the centre of the galaxy. This effect has been seen before in studies of the gas metallicity gradients in CALIFA galaxies by \citet{Rosales_2011}, \citet{Sanchez_2012b} and \citet{Sanchez_2014}.  \citet{Sanchez_2014}  studied the properties of 26 galaxies with clear evidence of a drop in the gas metallicity at their cores and a further 21 galaxies with possible evidence of a drop. Through this analysis they found no clear correlation with the morphology, presence of a bar, or through recent interactions or mergers, though they could not rule out the effect of a past interaction or minor merger that is no longer visible in the morphology of the galaxy. They did however find that many of the galaxies displaying this feature also appear to show a central star-forming ring, detected in the H$\alpha$ emission maps. They concluded that the drop in the gas metallicity in the center of the galaxy is associated with a central star-forming ring, and that  both features are produced by the radial flow of gas induced by resonances in the disk pattern speed. They further hypothesised that the apparent drop or flattening in the gas metallicity in the centres of galaxies could simply be an artefact of a ``hump'' in the gas metallicity at a radius of $\sim0.5~R_e$. \citet{SanchezM_2018} also found that the negative metallicity gradient across the discs of galaxies is often stronger  in galaxies with a drop in the core region than in those without the central drop. In the case of \Leda\ , the continuum-subtracted H$\alpha$ map, shown on the right of Fig.~\ref{fig:images}, doesn't show a clear ring of star formation, but there is a  drop in the H$\alpha$ emission in the central region of the galaxy, within a radius of $\sim 2\arcsec$. Furthermore, the WHAN and WHaD diagrams show that this region of the galaxy is retired, and the luminosity-weighted stellar populations in Fig~\ref{fig:radial_plots_stars} also show a sharp increase in the  ages within a radius of $2\arcsec$, further indicating a drop in the recent star formation in the central region of the galaxy. These trends (lower  $W_{\rm H\alpha}$ and older stellar ages) are consistent with the findings of \citet{Easeman_2022} for galaxies with drops in the central gas phase metallicity, and the origin of the lower gas phase metallicity is still an open question in the literature.

There are several possible explanations for the lower gas metallicity at the centre of the galaxy. For example, the galaxy could have accreted “pristine” or unenriched gas into its center, either from the disc or from an external source. Without ongoing star formation, this gas would not have been enriched in the same way as the rest of the disc, producing a dilution effect and leading to the positive metallicity gradient observed in the inner region  \citep[e.g.][]{Ceverino_2016, SanchezM_2018}. Alternatively, the more metal enriched gas could have been driven out of the galaxy in its history through past AGN activity or outflows driven by starbursts \citep{Easeman_2022}. Such events may have driven enough gas out of this region to quench star formation there (inside-out quenching), and as a result the enrichment of the ISM freezes while the rest of the disc becomes richer in metals \citep{Ceverino_2016, Belfiore_2017}.

\subsection{Star-Formation rate}\label{sec:SFR}
The final step in this part of the analysis was to measure the star-formation rate (SFR) across the galaxy. H$\alpha$ emission lines are one of the best SFR indicators in star forming galaxies observed in the optical since only young, massive stars with ages $<20~$Myr produce vast amounts of photons that ionise the surrounding gas \citep{Davies_2016}. Thus, the SFR calculated using H$\alpha$ is largely independent of star-formation history of the galaxy \citep{Kennicutt_1998}.

In order to derive accurate measurements for the SFR one must first consider dust extinction corrections. While the H$\alpha$ line is less sensitive to dust extinction compared to other star formation indicators, such as in the UV, dust effects can still be significant, even for local galaxies \citep{Koyama_2015}. We therefore used the dust corrected H$\alpha$ fluxes that we calculated in Section~\ref{sec:WHAN_WHAD} to calculate the H$\alpha$ luminosity, L(H$\alpha$), using
 \begin{equation}
L(H\alpha) = 4 \pi d_L^2F(H\alpha)_i,
\end{equation}
where $d_L$ is the luminosity distance, and then the SFR was determined using the relation
\begin{equation}
{\rm SFR}({H}\alpha)=5.5 \times 10^{-42} L({H}\alpha)
\end{equation}
from \citep{Calzetti_2013}. 
%This calculation is considered valid if star formation has remained constant over timescales $> 6$~Myr. 
%However, if there is contamination from an AGN, this calculation could overestimate the SFR in that region of the galaxy.

Having calculated the SFR for each bin, the total SFR for the galaxy as a whole was calculated to be SFR~$= 0.304 \pm 0.017 $~M$_\odot$, which is higher than the SFR calculated by \citet{Fan_2018} of $0.09 \substack{+0.07 \\ -0.02}$~M$_\odot$ using SED fitting. Generally, spectroscopic SFRs, especially those calculated using the H$\alpha$ line, are most sensitive to recent or ongoing star formation since only stars with masses $>10$~M$_\odot$ and lifetimes $<20$~Myr contribute significantly to the integrated ionizing flux \citep{Kennicutt_1998, Calzetti_2007}. Thus, the SFR derived from the H$\alpha$ emission traces stars with lifetimes of the order $3-10$~Myr \citet{Kennicutt_2012, Flores_2021}. On the other hand, SFRs calculated through SED fitting tend to be more sensitive to dust reddening and star formation over longer timescales, typically of the order $10-300$~Myrs, and so give a more time averaged SFR measurement \citep{Kennicutt_2012}.  Thus, the higher value measured in this work could indicate that the galaxy is undergoing a period of enhanced star formation compared to its recent history, but it is still on the low side for a typical spiral galaxy \citep[e.g.][]{Catalan_2015, Catalan_2017, Santoro_2022}. However, it should be noted that other factors may play a role in this discrepancy, such as aperture effects, dust treatment, IMF assumptions etc.
% Thus, the lower value measured in this work could indicate that the galaxy is undergoing quenching, where the SFR is declining with time. However, it should be noted that other factors may play a role in this discrepancy, such as aperture effects, dust treatment, IMF assumptions etc.

Finally, to account for the different areas covered by each bin across the galaxy, the SFR density, $\Sigma_{SFR}$, was calculated. The bottom row of Fig.~\ref{fig:radial_plots_gas} shows the $\Sigma_{SFR}$ as a function of radius from the center of the galaxy, and as a map on the right of the plot. In general, the $\Sigma_{SFR}$ is relatively flat across the disc, with a slight drop in the centre of the galaxy ($r\lesssim 1 \arcsec$) and in the outermost regions of the disc ($r\gtrsim 6 \arcsec$). 

For a typical star forming galaxy, the $\Sigma_{SFR}$ is generally $\sim 10^{-1}-10^{-3}$ M$_\odot$~yr$^{-1}$~kpc$^{-2}$ \citep{Leroy_2012, Cano_2016}, whereas Green-valley regions, which are partially quenched but still forming some stars, have typical $\Sigma_{SFR}\sim 10^{-3}-10^{-4}$ M$_\odot$~yr$^{-1}$~kpc$^{-2}$ \citep{Belfiore_2017}. Moving to lower values, cLIER bulges (retired region) in star-forming spiral galaxies have $\Sigma_{SFR}\sim 10^{-4}-10^{-5}$ M$_\odot$~yr$^{-1}$~kpc$^{-2}$ \citep{Belfiore_2017}, and fully retired galaxies with emission-line regions powered by HOLMES have $\Sigma_{SFR}\sim 10^{-5}$ M$_\odot$~yr$^{-1}$~kpc$^{-2}$ \citep{Gonzalez_2017}. Based on these values in the literature, it can be seen that \Leda\. appears to be a typical star-forming spiral, but with a relatively low $\Sigma_{SFR}$ that approaches the  values normally seen in green valley galaxies. Thus, these results may be pointing towards a scenario where \Leda\. is in the early stages of being quenched, possibly from the inside out.

%%%%%%%%%%%%%%%%%%%%%%%%%%%%%%%%%%%%%%%%%%%%%%%%%%%%%%%%%%%%%%
\section{Summary and Conclusions}\label{sec:summary}
In this work we present a detailed study of \Leda\. , a flocculent spiral galaxy that was found in the foreground of a deep MUSE datacube with a $\sim20$ hour exposure time. While many observations of spiral galaxies at this redshift ($z\sim0.009$) exist in surveys like MaNGA, CALIFA and SAMI, they lack the spatial resolution and the full coverage of the galaxy that we have in this datacube, meriting a more detailed study of its stellar and gas properties over the full extent of the galaxy.

A study of the visual morphology and the stellar and gas kinematics show no obvious signs of distortion, indicating that it is unlikely that the galaxy has undergone a recent interaction within the last 1 billion years \citep{Hung_2016, Davis_2016}. In the datacube, there does appear to be a neighbouring galaxy $\sim20\arcsec$ to the north-west of \Leda\., but analysis of their redshifts and distances revealed that the two galaxies have a physical separation of $\sim1$~Mpc, meaning that while they may be part of the same group or cluster of galaxies, they are too far apart to be directly interacting.

Analysis of the stellar populations showed relatively old mass-weighted ages across the whole galaxy, with older ages int he bulge-dominated region at the center and no strong gradient across the disc, indicating that the galaxy formed the majority of its mass long ago.  On the other hand, the luminosity-weighted ages show a steep negative trend in the inner $2-3\arcsec$ with a relatively flat distribution of young stars across the rest of the disc. The stellar metallicities, both luminosity- and mass-weighted, show a negative gradient with radius across the disc, which is often seen in spiral galaxies. In the inner part of the galaxy, the mass-weighted metallicities show a slight positive gradient before decreasing again throughout the rest of the disc.  This drop at the very centre coinicdes with the region of older luminosity- weighted stellar populations, and is consistent with the scenario where in that region the star formation has been truncated, reducing the chemical enrichment there relative to other parts of the disc that have ongoing star formation. In the outer part of the disc the luminosity-weighted metallicities are slightly higher than the mass-weighted values, indicating stronger chemical enrichment in the younger, more recently formed stars than the older stars. 

Analysis of the gas properties revealed that the gas metallicities also show a negative gradient outside of a radius of $2-3\arcsec$, but display a positive gradient within this radius. The $\Sigma_{SFR}$ is also relatively flat, with a slight drop in the innermost region of the galaxy. The BPT diagram was also created, which indicated that the majority of the disc is star forming, but the central region lay in the LINER part of the diagram. A deeper analysis of the properties of the H$\alpha$ emission lines using the WHAN and WHaD diagrams, which also consider the equivalent width and velocity dispersion of the H$\alpha$ line, revealed that actually the centre of the galaxy is a "retired" region, where the main source of ionization is not from a LINER, but from hot, old, massive evolved stars (HOLMES).

Together, these results reflect that \Leda\. is a relatively normal spiral galaxy. The results indicate that the majority of its mass was created a long time ago, likely around $6-7$~billion years ago based on the mass-weighted ages across the galaxy, and that the majority of the disc has undergone continuous star formation since then. 

The central region of the galaxy, within a radius of $\sim2\arcsec$, is interesting though. The luminosity-weighted ages show a sharp increase in this region, indicating that the light is dominated by older stars there. These stars could come from the bulge, which is typically older than the disc \citep{Lah_2023, Jin_2024, Jegatheesan_2024}, but could also reflect truncation or quenching of star formation in the centre of the galaxy. This scenario is also consistent with the lower $\Sigma_{SFR}$ at the center of the galaxy and the "retired region" classification from the WHAN and WHaD diagrams. The $\Sigma_{SFR}$ is actually on the lower end of the range for typical star forming galaxies, touching on values often seen in green valley galaxies, so it is possible that this galaxy is in the early stages of inside-out quenching.

The gas metallicity plot is also consistent with this scenario. The negative gradient in the outer disc can be explained by lower-metallicity gas being accreted into the outskirts of the galaxy and the gas generally becoming more metal-enriched by ongoing star formation as it migrates inwards through the disc. The drop in the gas metallicity in the centre could then be explained by the star formation there being slowly quenched or truncated, leading to less enrichment over time than in the regions surrounding it.

In summary, these results indicate that \Leda\. is a nice example of a cLIER galaxy \citep{Belfiore_2016}, where the central kpc or so is dominated by old, metal-poor stars with little star formation and where the central LIER emission is primarily powered by hot evolved (pAGB) stars, while the rest of the disc is displaying ongoing star formation. Often these galaxies are found to show evidence of inside-out quenching, such as seen in \Leda\ . The unusually deep observations with MUSE of this galaxy give us a rare opportunity to study this phenomenon with better spatial resolution than can normally be achieved with the current IFU surveys of galaxies at a similar redshift.

\begin{acknowledgements}
We would like to thank the anonymous referee for their comments, which helped to improve the paper.
This work was based on observations collected at the European Organisation for Astronomical Research in the Southern Hemisphere under ESO programmes 106.21F0.001 120
(PI Diaz-Santos) and 109.2393.001 (PI Assef).
This research has made use of the NASA/IPAC Infrared Science Archive, which is funded by the National Aeronautics and Space Administration and operated by the California Institute of Technology.
E.J.J, S.L., R.J.A, M.A. and D.S. acknowledge the support from the ANID CATA-BASAL project FB210003.
E.J.J was supported by FONDECYT Regular grant number 1262304.
A.Z.L.A. gratefully acknowledges the support provided by the Postdoctoral Program (POSDOC) of UNAM (Universidad Nacional Autónoma de México). 
M.A. is supported by FONDECYT grant number 1252054, and gratefully acknowledges support from ANID MILENIO NCN2024\_112 and ANID + Vinculaci\'on Internacional + FOVI250261. 
M.L. was supported by the grants from the Rubin-Chile Fund (DIA3324).
D.S. was supported by the ALMA-ANID grant number 31220030.
R.J.A was supported by FONDECYT grant number 1231718.
%\\

\end{acknowledgements}

\bibliographystyle{aa} % style aa.bst
\bibliography{references} % your references Yourfile.bib

\end{document}